\documentclass[a4paper,11pt]{article}
\pdfoutput=1 

\usepackage{jheppub} 

\usepackage{amsmath,amssymb,bm,bbm}
\usepackage[utf8]{inputenc}

\title{\boldmath Rectangular W-algebras, extended higher spin gravity and dual coset CFTs}

\author[a]{Thomas Creutzig}
\author[b]{Yasuaki Hikida}

\affiliation[a]{Department of Mathematical and Statistical Sciences, University of Alberta,\\Edmonton, Alberta T6G 2G1, Canada}

\affiliation[b]{Center for Gravitational Physics, Yukawa Institute for Theoretical Physics, Kyoto University,\\Kyoto 606-8502, Japan}

\emailAdd{creutzig@ualberta.ca}
\emailAdd{yhikida@yukawa.kyoto-u.ac.jp}

\abstract{We analyze the asymptotic symmetry of higher spin gravity with $M \times M$ matrix valued fields, which is given by rectangular W-algebras with su$(M)$ symmetry. The matrix valued extension is expected to be useful for the relation between higher spin gravity and string theory. With the truncation of spin as $s=2,3,\ldots , n$, we evaluate the central charge $c$ of the algebra and the level $k$ of the affine currents with finite $c,k$.  For the simplest case with $n=2$, we obtain the operator product expansions among generators by requiring their associativity.  We conjecture that the symmetry is the same as that of Grassmannian-like coset based on our proposal of higher spin holography. Comparing $c,k$ from the both theories, we obtain the map of parameters. We explicitly construct low spin generators from the coset theory, and, in particular, we reproduce the operator product expansions of the rectangular W-algebra for $n=2$. We interpret the map of parameters by decomposing the algebra in the coset description.
}

\keywords{Conformal and W Symmetry, AdS-CFT Correspondence, Higher Spin Gravity}
\arxivnumber{1812.07149}
\preprint{YITP-18-129}

\begin{document}
	\maketitle
	\flushbottom

\section{Introduction}

Higher spin symmetry is believed to be useful to investigate the tensionless limit of string theory \cite{Gross:1988ue}.
A higher spin gauge theory on AdS space is given by a Vasiliev theory  \cite{Vasiliev:2003ev} with a gauge field for each spin $s=2,3,\ldots$, 
and the theory is expected to describe the first Regge trajectory of strings.
In order to explain the higher Regge trajectories as well, we may consider the Vasiliev theories with $M \times M$ matrix valued fields, see \cite{Vasiliev:2018zer} for higher tensor generalizations. 
It was proposed in \cite{Chang:2012kt} that the matrix extension of 4d Vasiliev theory is dual to the Aharony-Bergman-Jafferis(-Maldacena) theory \cite{Aharony:2008ug,Aharony:2008gk}, and the duality implies a connection between the higher spin gravity and superstrings on AdS$_4 \times \mathbb{CP}^3$.
In our previous work \cite{Creutzig:2013tja},
we examined  the 3d Prokushkin-Vasiliev theory with matrix valued fields \cite{Prokushkin:1998bq}, and we claimed that the theory is dual to a 2d coset model.%
\footnote{Several generalizations of the holographic duality were discussed in \cite{Eberhardt:2018plx}.}
In this paper, we extend the analysis on the asymptotic symmetry of the 3d higher spin gravity beyond the classical limit.
The symmetry is infinite dimensional and should be useful to analyze quantum corrections.
In  \cite{Creutzig:2013tja}, we have considered a bosonic and a $\mathcal{N}=2$ supersymmetric holography. 
Without the matrix extension, i.e., with $M =1$, the bosonic holography essentially reduces to the one by \cite{Gaberdiel:2010pz}.
Moreover, the  $\mathcal{N}=2$ holography with $M=1$ is identical to that of \cite{Creutzig:2011fe}.
In this paper, we mainly focus on the bosonic case.

For $M=1$, the gauge algebra of the 3d higher spin theory is called as hs$[\lambda]$, which can be truncated to sl$(n)$ for $\lambda = n$ $(n = 2,3, \ldots)$. 
After the matrix extension, the gauge algebra becomes
\begin{align}
\text{sl}(M) \otimes  \mathbbm{1}_n  \oplus \mathbbm{1}_M \otimes  \text{sl}(n) \oplus  \text{sl}(M) \otimes  \text{sl}(n) \simeq \text{sl}(M n) \, ,
\label{slMn}
\end{align}
see, e.g., \cite{Creutzig:2013tja,Joung:2017hsi}.
The gravitational sector corresponds to $\mathbbm{1}_M \otimes \text{sl}(2) \subset \mathbbm{1}_M \otimes \text{sl}(n) $ with the principal embedding of $\text{sl}(2)$ into $\text{sl}(n)$.%
\footnote{Different embeddings of $\text{sl}(2)$ were analyzed in \cite{Gwak:2015vfb,Gwak:2015jdo}.}
The sl$(2)$ embedding  corresponds to the partition of $M n$;
\begin{align}
M n = n + n + \cdots + n \, .
\label{partition}
\end{align}
The asymptotic symmetry of higher spin gravity has been analyzed in \cite{Henneaux:2010xg,Campoleoni:2010zq,Gaberdiel:2011wb,Campoleoni:2011hg}.
Utilizing the arguments, the asymptotic symmetry can be obtained as the rectangular W-algebra\footnote{The partition \eqref{partition} corresponds to the rectangular Young tableau consisting of $M$ columns of $n$ boxes of height each, hence the name rectangular W-algebra.} from the Hamiltonian reduction of $\text{sl}(M n)$ with the sl$(2)$ embedding \eqref{partition}.
We can see that the affine $\text{su}(M)$ Lie algebra is included as a sub-algebra.
We first compute the central charge $c$ of the W-algebra and the level $k$ of the affine $\text{su}(M)$ at the classical limit, and then obtain the exact results of $c,k$  by applying the quantum Hamiltonian reduction. We further determine the operator product expansions (OPEs) of generators for $n=2$ by assigning their associativity and show the uniqueness of the algebra, see \cite{Joung:2017hsi} for the analysis at the classical limit.

In our previous work \cite{Creutzig:2013tja}, we have proposed that the 3d Vasiliev-Prokushkin theory with $M \times M$ matrix valued fields is dual to the Grassmannian-like coset
\begin{align}
\frac{\text{su}(N+M)_k}{\text{su}(N)_k \oplus \text{u}(1)_{kNM(N+M)}} \, .
\label{Grassmann}
\end{align}
The classical higher spin theory corresponds to the large $N$ limit with the 't Hooft parameter%
\footnote{The definition of 't Hooft parameter in \cite{Creutzig:2013tja} is slightly different from the one here at finite $k,N$, though they are the same at the large $N$ limit. The difference is important since we would like to analyze with finite $k,N$.}
\begin{align}
\lambda = \frac{k}{k + N}
\label{tHooft}
\end{align}
and $M$ kept finite. 
The parameter is identified with  $\lambda$ for the higher spin algebra hs$[\lambda]$.
As a strong support for our conjecture, we have shown the match of one-loop partition functions at the 't Hooft limit among others.
Based on this duality, we propose here that the W-algebra obtained from the Hamiltonian reduction of sl$(M n)$ is the same as the one for the coset \eqref{Grassmann} with $\lambda = n$.
This leads to $k = - n N / ( n - 1 )$, which implies that the coset theory is in a non-unitary regime.
We provide the map of parameters by comparing the central charge of the algebra and the level of affine $\text{su}(M)$.
We explicitly construct low spin currents transforming in the adjoint representation of su$(M)$,
and reproduce the OPEs for the rectangular W-algebra with $n=2$.
We can use another 't Hooft parameter
\begin{align}
\lambda = - \frac{k}{k + N + M}
\label{tHooftp}
\end{align}
in stead of \eqref{tHooft}, and this implies a duality relation of the coset model \eqref{Grassmann}
as in the cases with $M=1$ \cite{Gaberdiel:2012ku,Prochazka:2014gqa,Candu:2012tr}.
We give an explanation of the duality by decomposing the W-algebra in terms of the coset description as in \cite{Gaiotto:2017euk,Creutzig:2017uxh,Prochazka:2017qum,Prochazka:2018tlo,Harada:2018bkb}.

\subsection{The main conjecture}

The main conjecture of this work is that the coset \eqref{Grassmann} at level $k$ is isomorphic to the simple rectangular $W$-algebra of sl$(Mn)$ at level $-t$ where the levels are related via
$k = - t n+Mn(n-1)$, if 
\begin{align}
 k =- \frac{n N}{n - 1} \qquad\text{or} \qquad k  = - \frac{n (N + M)}{n+1}  \, .
\end{align}
In the first case the first 't Hooft parameter \eqref{tHooft} is
\begin{align}
\lambda = \frac{k}{k + N} = n
\end{align}
while in the second case the second  't Hooft parameter \eqref{tHooftp} is
\begin{align}
\lambda = - \frac{k}{k + N + M} =n
\end{align}
as well. 
We expect it to be a very difficult problem to give a general proof of this conjecture. We verify that central charges of theories and levels of current algebras agree. Furthermore in the second case $ k  = - n (N + M)/( n+1 ) $ we observe from the characters that the coset indeed has the required null vectors at conformal weight $n+1$. We then prove a uniqueness result of this type of W-algebra for the case $n=2$, and various $N, M$. This means that a simple chiral algebra of this type with strong generators only in weight one and two is completely determined by the level of the current algebra and the central charge. 
We then also check that indeed there are no fields of higher spin for various cosets at $\lambda=2$ so that we have proven our conjecture in these cases.

\subsection{Organization}

The organization of this paper is as follows;
In the next section, we introduce the higher spin gravity with $M \times M$ matrix valued fields and the rectangular W-algebra as its asymptotic symmetry near the AdS boundary.
For this, we simply apply the general procedure in \cite{Henneaux:2010xg,Campoleoni:2010zq,Gaberdiel:2011wb,Campoleoni:2011hg}. In particular, we compute the central charge $c$ of the W-algebra and  the level $k$ of su$(M)$ currents at the classical limit as was done in \cite{Creutzig:2013tja} for the $\mathcal{N}=2$ supersymmetry setup.
In section \ref{sec:quantumsym}, we start by providing some overview of W-algebras.
We then propose the actions for the symmetry algebras. We obtain the exact expressions of  $c,k$ both from the general prescription and the proposed actions, and we find agreement.  
For the simplest case with $n=2$, we compute the OPEs of generators using their associativity.
In section \ref{sec:Grassmannian}, we realize the W-algebra in terms of coset description based on our proposed holography in \cite{Creutzig:2013tja}. Comparing $c,k$, we obtain the map of parameters between the algebras from the reduction of sl$(M n)$ and the coset model \eqref{Grassmann}.
We then explicitly construct spin 2 currents transforming in the adjoint representation of su$(M)$
and reproduce the OPEs for $n=2$.
Decomposing the W-algebra in terms of coset model, we give an interpretation of the map of parameters.
Section \ref{sec:conclusion} is devoted to the summary of this paper and discussions on open problems.
In appendix \ref{sec:spin3}, we construct spin 3 currents of the W-algebra from the coset model \eqref{Grassmann}.
In appendix \ref{sec:webofwalg}, we reproduce the decomposition of the W-algebra in terms of the algebra obtained from the reduction of sl$(M n)$ for the simplest setup with $n=2$ and $M=2$.

\section{Asymptotic symmetry of higher spin gravity}
\label{sec:Vasiliev}

In this section, we examine the classical asymptotic symmetry of higher spin gravity near the AdS boundary.
Higher spin gauge theory can be constructed from  Chern-Simons gauge theory based on $g \oplus g$ Lie algebra. The action is given by
\begin{align}
S = S_\text{CS} [A] - S_\text{CS} [\tilde A] \, , \quad
S_\text{CS} [A ] = \frac{k_\text{CS}}{4 \pi}\int  \text{Tr} \left( A \wedge d A + \frac{2}{3} A \wedge A \wedge A \right) \, ,
\label{action}
\end{align}
which is invariant under the gauge transformations
\begin{align}
\delta A = d \Lambda + [A , \Lambda] \, , \quad \delta\tilde  A = d \tilde \Lambda + [\tilde A , \tilde \Lambda] \, .
\label{gaugetrans}
\end{align}
The 1-from gauge fields $A , \tilde A$ and the gauge parameters $\Lambda , \tilde \Lambda$ take values in $g$.
The theory describes pure AdS$_3$ gravity for $g = \text{sl}(2)$ \cite{Achucarro:1987vz,Witten:1988hc} and higher spin gravity for a higher rank $g$ \cite{Blencowe:1988gj}. 
In the next subsection, we introduce a higher spin algebra  $g$ relevant for the higher spin gravity with the matrix extension.
In subsection \ref{sec:asymptotic}, we examine the asymptotic symmetry of the higher spin gravity at the classical limit applying the method of  \cite{Henneaux:2010xg,Campoleoni:2010zq,Gaberdiel:2011wb,Campoleoni:2011hg}.

\subsection{Higher spin algebra}

We start from the 3d Prokushkin-Vasiliev theory without the matrix extension in \cite{Prokushkin:1998bq}.
The gauge algebra is given by a higher spin algebra $\text{hs} [\lambda]$.
One definition is 
\begin{align}
\text{hs}[\lambda] = B[\lambda] \ominus \mathbbm{1} \, ,
\end{align}
where
$B[\lambda]$ is the universal enveloping algebra of sl(2) divided by the ideal $C_2 - \frac14( \lambda^2 -1) \mathbbm{1}$ ($C_2$ is the sl(2) Casimir);
\begin{align}
B[\lambda] = \frac{U(\text{sl}(2))}{ \langle C_2 - \frac14( \lambda^2 -1) \mathbbm{1} \rangle} \, .
\end{align}
At $\lambda = n$ with $n=2,3,\ldots$, an ideal appears and $\text{hs} [\lambda]$ can be truncated to sl$(n)$ by dividing the ideal.
We may express the generators of $B[\lambda]$ as $V^1_0 \equiv \mathbbm{1}$ and $V^s_m$ with $s=2,\ldots$ and $m = -s + 1 , \ldots , s-1$, where  the commutation relations are of the form
\begin{align}
[V_m^s , V_l^t] = \sum_{u=2,4,\ldots}^{s + t - |s - t|-1} g_u^{st} (m,l; \lambda) V_{m+l}^{s + t -u} \, .
\end{align}
We use the definition of trace;
\begin{align}
\begin{aligned}
&\text{tr}_V (V^s_m V^t_l) = N_s \frac{(-1)^{s-m-1}}{(2 s - 2)!} \Gamma(s + m) \Gamma (s - m) \delta^{st} \delta_{m , - l} \, ,  \\
&
N_s = \frac{3 \cdot 4^{1 -s } \sqrt{\pi}  \Gamma (s) \Gamma (s - \lambda) \Gamma (s + \lambda)}{(\lambda^2 - 1) \Gamma(s + \frac12) \Gamma (1 - \lambda) \Gamma (1 + \lambda)}  \, .
\end{aligned}
\label{defoftrace}
\end{align}
In particular, we need
\begin{align}
\text{tr}_V (V^1_0 V^1_0) = - \frac{6}{1 - \lambda ^2 } \, , \quad
\text{tr}_V (V^2_{0} V^2_0) = \frac{1}{2} \, .
\label{traceVV}
\end{align}
See \cite{Pope:1989sr,Gaberdiel:2011wb} for more details of the algebra.

We extend the algebra by multiplying the $M \times M$ matrix algebra as (see, e.g., \cite{Gaberdiel:2013vva,Creutzig:2013tja})
\begin{align}
\text{hs}_M [\lambda] = \text{gl}(M) \otimes B[\lambda] \ominus  \mathbbm{1}_M \otimes \mathbbm{1} \, .
\label{hsMlambda}
\end{align}
This algebra can be decomposed as
\begin{align}
\text{hs}_M [\lambda]  =\text{sl}(M) \otimes \mathbbm{1} \oplus \mathbbm{1}_M \otimes \text{hs}[\lambda]  \oplus \text{sl}(M) \otimes \text{hs}[\lambda] \, .
\label{decomposition}
\end{align}
Notice that only the identity operator $ \mathbbm{1}_M \otimes \mathbbm{1}$ can be decoupled from the algebra.
There are two sub-algebras $\text{sl}(M) \otimes \mathbbm{1} $ and $\mathbbm{1}_M \otimes \text{hs}[\lambda]$, and the sub-sectors with these sub-algebras do not mix with the others by the gauge transformations in \eqref{gaugetrans}.
We denote the generators of $\text{gl}(M) $ as $t^A = \mathbbm{1}_M , t^a$ $(a = 1, \ldots ,M^2-1)$ with the normalization
\begin{align}
\text{tr}_M (t^a t^b) = \delta^{ab} \, , \quad [t^a , t^b] = i f^{ab}_{~~c} t^c \, .
\label{suMgenerators}
\end{align}
Here we express $t^A$ by a $M \times M$ matrix and define the trace  tr$_M$ as the sum over the diagonal elements.
The generators of $\text{hs}_M [\lambda]  $ are given by $t^A \otimes V_m^s$, and the trace is defined as the product
\begin{align}
\text{Tr}  [( t^A \otimes V_m^s) (t^B \otimes V_l^t) ] =  \text{tr}_M ( t^A t^B )  \text{tr}_V ( V_m^s V_l^t ) \, .
\label{tracefull}
\end{align}

In the next subsection, we examine the asymptotic symmetry at the classical limit, and in the following sections, we extend the analysis including quantum effects. In order to deal with quantum corrections, it is convenient to work with $\lambda =n$.
The algebra after the matrix extension is then given by sl$(Mn)$ as in \eqref{slMn}.
We may express the generators of the algebra as $t^A \otimes  V^s_m$ as before but with a restriction $s \leq n$.
For the generators of gl$(n)$, we can use $n \times n$ matrices, see, e.g., \cite{Castro:2011iw} for the explicit realizations.
The trace $\text{tr}_V = \text{tr}_n$ is given as in the gl$(M)$ sector, and the current convention leads to
\begin{align}
\text{tr}_V (V^1_0 V^1_0) = \text{tr}_n (\mathbbm{1}_n)  = n  \, , \quad
\text{tr}_V (V^2_{0} V^2_0) = \frac{1}{12} n (n^2 - 1)
\label{traceVV0}
\end{align}
for example.

\subsection{Classical asymptotic symmetry}
\label{sec:asymptotic}

In order to define a higher spin gravity, we need to assign proper boundary conditions to higher spin fields.
For the purpose, we introduce the coordinate system $(\rho , z , \bar z)$, where
$\rho$ is the radial coordinate and the boundary is located at $\rho \to \infty$.
Moreover, $(z , \bar z)$ are the coordinates for the plane parallel to the boundary. 
The gravitational sector is identified as 
$
\mathbbm{1}_M \otimes \text{sl}(2) \subset \mathbbm{1}_M \otimes \text{hs}[\lambda] 
$
with the principally embedded $\text{sl}(2) \subset \text{hs}[\lambda]$.
The solutions to the equations of motion from the Chern-Simons theory are given by flat connections, which may be expressed by
\begin{align}
A = e^{- \rho ( \mathbbm{1}_M \otimes V_0^2 )} a (z)   e^{\rho(  \mathbbm{1}_M \otimes V_0^2 )} d z +( \mathbbm{1}_M \otimes  V_0^2 ) d \rho 
\label{A}
\end{align}
in a gauge fixing and similarly for $\tilde A$.
The AdS background corresponds to the configuration of gauge field $A_\text{AdS}$ with $a(z) = \mathbbm{1}_M \otimes V_1^2$.
We assign the boundary condition such that \cite{Henneaux:2010xg,Campoleoni:2010zq}
\begin{align}
\left. (A - A_\text{AdS}) \right|_{\rho \to \infty} = \mathcal{O} (1) \, .
\end{align}
With this condition and the residual gauge symmetry, we can put
\begin{align}
\begin{aligned}
a(z) =   \mathcal{J}_a (z) ( t^a \otimes V_0^1 ) + \mathbbm{1}_M \otimes V_1^2 & +   \sum_{s = 2}^\infty \mathcal{W}^{(s)} (z) ( \mathbbm{1}_M \otimes V_{-s+1}^s ) 
 \\ &  + \sum_{s = 2}^\infty \mathcal{Q}^{(s)}_a (z) (t^a \otimes V_{-s+1}^s ) \, ,
\end{aligned}
\label{hwg}
\end{align} 
see  \cite{Balog:1990mu,Campoleoni:2017xyl,Joung:2017hsi}.

The asymptotic symmetry can be obtained as a classical Hamiltonian reduction of $\text{hs}_M [\lambda]$ for the Chern-Simons theory by following the standard procedure in \cite{Henneaux:2010xg,Campoleoni:2010zq,Gaberdiel:2011wb,Campoleoni:2011hg}.
We can read off the algebra generated by $\mathcal{J}_a,\mathcal{W}^{(s)},\mathcal{Q}^{(s)}_a$ from the gauge transformations in \eqref{gaugetrans} preserving the form of gauge fields as in \eqref{hwg}.
We can see that $\mathcal{J}_a$ generates  affine sl$(M)$ (or su$(M)$) Lie algebra.
Moreover,  $\mathcal{W}^{(s)}$ and $\mathcal{Q}^{(s)}_a$ are spin $s$ currents transforming in the trivial and adjoint representation of sl$(M)$, respectively.
It is quite complicated to obtain the full commutation relations among the generators, see \cite{Joung:2017hsi} for the simplest case with $\lambda =2$.
Thus, in the following, we focus on the two sub-sectors including $\mathcal{J}_a$ and $\mathcal{W}^{(s)}$, and compute the level of affine sl$(M)$ and the central charge of the algebra.
Similar analysis has already been done in \cite{Creutzig:2013tja} for the $\mathcal{N}=2$ supersymmetric case.

First, we restrict ourselves to the sub-sector with $\mathcal{J}_a$.
We set the other elements  $\mathcal{W}^{(s)},\mathcal{Q}^{(s)}_a$ to be zero and consider the gauge transformations \eqref{gaugetrans} with the generators $t^a \otimes V^1_0$. 
With this setup, the system becomes the same as the sl$(M)$ Chern-Simons gauge theory, so the boundary symmetry is the affine sl$(M)$ Lie algebra. 
From the definition of trace in \eqref{tracefull}, we have
\begin{align}
\text{Tr} [( t^a \otimes V^1_0 )( t^b \otimes V^1_0)] 
= \text{tr}_M ( t^a t^b ) \text{tr}_V (V^1_0 V^1_0) \, .
\end{align}
This means that the overall factor in front of the action is multiplied by $\text{tr}_V (V^1_0 V^1_0)$.
Therefore, the level of affine sl$(M)$ can be read off as
\begin{align}
\ell = k_\text{CS} \text{tr}_V (V^1_0 V^1_0) =  - \frac{6 k_\text{CS}}{1 - \lambda^2 }\, .
\label{slMlevel}
\end{align}
Here we have used \eqref{traceVV}. For the comparison with the symmetry of the coset \eqref{Grassmann}, it is convenient to 
use the term of su$(M)$ instead of sl$(M)$. The level $k$ of affine su$(M)$ is related as $ k = - \ell$, thus we have
\begin{align}
k = \frac{6 k_\text{CS}}{1 - \lambda^2 }
\label{slMlevelk}
\end{align}
instead of \eqref{slMlevel}.

Next, we restrict ourselves to the sub-sector with $\mathcal{W}^{(s)} $ in a similar way.
We set the other elements  $\mathcal{J}_a,\mathcal{Q}^{(s)}_a$ to be zero and consider the gauge transformations \eqref{gaugetrans} with the generators $\mathbbm{1}_M \otimes V^s_m$. 
Then the setup is the same as the hs$[\lambda]$ Chern-Simons theory with the principal embedding of sl(2) in hs$[\lambda]$. The W-algebra obtained this way is called as W$_\infty[\lambda]$ but now at the classical limit.
The definition of the trace in \eqref{tracefull} leads to
\begin{align}
\text{Tr} [( \mathbbm{1}_M \otimes V^s_m )( \mathbbm{1}_M \otimes V^t_l) ]
= \text{tr}_M ( \mathbbm{1}_M ) \text{tr}_V (V^s_m V^t_l)  = M \text{tr}_V (V^s_m V^t_l) \, .
\end{align}
Therefore, the central charge of the algebra $W_{\infty}[\lambda]$ is \cite{Henneaux:2010xg,Campoleoni:2010zq}
\begin{align}
c = 12  k_\text{CS} M  \text{tr}_V (V^2_0 V^2_0) = 6 k_\text{CS} M 
\label{centerCS}
\end{align}
using \eqref{traceVV}.
Since $\mathcal{W}^{(2)} $ is the only spin 2 generator in the singlet of sl$(M)$, 
the central charge is the one for the whole algebra as well.
Removing $k_\text{CS}$, we find the relation
\begin{align}
c   =   M (1 - \lambda^2 ) k
\label{c2l}
\end{align}
from \eqref{slMlevelk} and \eqref{centerCS}.
In particular, the level $k$ is positive when $c > 0$ and $0 < \lambda <1$ as in the setups of the holography \cite{Gaberdiel:2010pz,Creutzig:2013tja}.

Since the two sub-sectors decouples, the affine su$(M)$ generators and the energy-momentum tensor commute with each other. Therefore, we should redefine the energy-momentum tensor for the affine su$(M)$ generators to be Virasoro primary as in \cite{Henneaux:1999ib}.
Let us denote the affine su$(M)$ generators as $\mathcal{J}^a_m$ and the Virasoro generators as ${\mathcal{L}}_m$.
We may choose the normalizations such that their commutation relations become
\begin{align}
\begin{aligned}
&i \{ {\mathcal{L}}_m , \mathcal{L}_l \} = (m - l) {\mathcal{L}}_{m+l} + \frac{c}{12} m (m^2 -1) \delta_{m+l,0} \, , \\
&i \{ \mathcal{J}^a_m , \mathcal{J}^b_l \} = k \delta^{ab} m \delta_{m+l,0} +  i f^{ab}_{~~c} \mathcal{J}^c_{m+l} \, , \quad i \{ {\mathcal{L}}_m , \mathcal{J}^a_l \} = 0 \, . \label{cmbefore}
\end{aligned}
\end{align}
Here the structure constant $f^{ab}_{~~c}$ is given as in \eqref{suMgenerators}.
Redefining the Virasoro generators as%
\footnote{With the notation in \eqref{suMgenerators}, we use $\delta^{ab}$ to raise or lower the su$(M)$ indices.}
\begin{align}
{\mathcal{L}}_m + \frac{1}{2 k} \sum_{l} J^a_{l} J^a_{m - l} \to \mathcal{L}_m \, ,
\end{align}
we obtain the commutation relations in \eqref{cmbefore} but the third one is replaced by
\begin{align}
i \{ {\mathcal{L}}_m , \mathcal{J}^a_l \}  = - l \mathcal{J}^a_{m + l}
\end{align}
for large $c,k$.

From the next sections, we mainly focus on the case with $\lambda = n$.
The relation between $c$ and $k$ is given by \eqref{c2l} with setting $\lambda = n$ as
\begin{align}
c   =  - M (n^2 - 1) k \, .
\label{c2l2}
\end{align}
For $c > 0$ and $n=2,3,\ldots$, the level $k$ become negative.
It might be worth to derive the same result with the conventional notation in terms of sl$(n)$ instead of hs$[\lambda]$.
The sub-sector with $\mathcal{J}_a $ leads to the affine su$(M)$ symmetry with the level
\begin{align}
k = - k_\text{CS} n  \, ,
\end{align} 
where we have used \eqref{slMlevel} with $k = - \ell$ and \eqref{traceVV0}.
The sub-sector with  $\mathcal{W}^{(s)}$ provides the $W_n$-algebra, which can be regarded as a truncation of W$_\infty[\lambda]$ with $\lambda = n$. 
The central charge of the algebra is 
\begin{align}
 c =  k_\text{CS} M n (n^2 -1) 
\end{align} 
using \eqref{centerCS} and \eqref{traceVV0}.
Using $k=-k_\text{CS}n$, we reproduce \eqref{c2l2}.

\section{Quantum Hamiltonian reduction}
\label{sec:quantumsym}

In the previous section, we have examined the asymptotic symmetry of the higher spin gravity at the classical limit.
In the case with $\lambda = n$, the asymptotic symmetry is given by the classical Hamiltonian reduction of sl$(Mn)$ with the sl(2) embedding corresponding to the partition \eqref{partition}. 
In this section, we study the W-algebra obtained as the quantum Hamiltonian reduction instead of the classical one, see \cite{Arakawa:2016rwm} for an introduction. 
Before we start, let us give some short overview about W-algebras. 

By a W-algebra we mean the chiral algebra or vertex algebra of a CFT that has generating fields of higher conformal weight, i.e. not only a current algebra. 
There are three standard constructions of such algebras, namely as BRST-cohomology, as joint intersection of kernels of screening operators and as coset.
Each approach has its advantages and difficulties. In the cohomological approach it is easy to determine the generating fields of the algebra and their spin, the kernel of screening picture gives a concrete action of the theory and is most suitable for concrete computations while cosets are often quite useful for understanding the representation theory. The spin content of coset theories can be determined using the theory of \cite{Creutzig:2012sf,Creutzig:2014lsa}.
The most common BRST-cohomology is the quantum Hamiltonian reduction, see \cite{Kac:2003jh}, and recently Genra  \cite{Genra:2016xxc} has shown that there is also a kernel of screening realization of these W-algebras. The most common W-algebras are the principal W-algebras corresponding to the principal or regular quantum Hamiltonian reduction. The ADE-series of these W-algebras is now finally also known to be realized as a coset theory \cite{Arakawa:2018iyk}. This is important as the coset realization of the $A$-series is precisely the dual theory for the ordinary bosonic higher spin gravity correspondence of Gaberdiel and Gopakumar \cite{Gaberdiel:2010pz}. Moreover the proof of \cite{Arakawa:2018iyk} consists of finding a kernel of screenings realization of the coset theory and this step generalizes and some generalizations are currently work in progress. 

Now, the bosonic higher spin algebra of  Gaberdiel and Gopakumar is of type $2, 3, 4, \dots$ and Linshaw has proven that there exists a two parameter family of such W-algebras \cite{Linshaw:2017tvv} and there are certain curves (in the parameter space) of ideals where the simple quotient truncates to an algebra of type $2, 3, \dots, n$ for some given $n$. Intersections of such curves correspond to isomorphisms of algebras and this is quite important to us.
A similar Theorem in the even spin case also exists \cite{Kanade:2018qut}. Let us now turn to W-algebras with su$(M)$ symmetry. By this we mean a vertex algebra that has a su$(M)$ current algebra together with $M^2$ generating fields of spin $2, 3, \dots, n$  for some $n$. These algebras will be parameterized by their field content, the central charge, the level of the current algebra and probably further parameters will be necessary for larger $n$. 
There are two natural ways to realize such W-algebras, namely rectangular W-algebras via quantum Hamiltonian reduction and the coset algebras of Grassmannian type cosets. We believe that there exists a W$_\infty(m)$-algebra, that is a multi parameter family of vertex algebras with an su$(M)$ current algebra and $M^2$ generators of spin $2, 3, \dots$ up to infinity. Further in analogy to W$_\infty$ there should be certain curves of ideals for which the simple quotient of the algebra truncates to an algebra with $M^2$ fields of spin $2, 3, \dots, n$ together with the su$(M)$ current algebra. Moreover the expectation is that both the Grassmannian type coset as well as the rectangular W-algebras are described by such quotients and moreover we expect coincidences, i.e. intersections of these curves and interesting values of the parameters. In the following we will explore this idea.

We start by introducing rectangular W-algebras and proposing an action for these theories. Especially we compute the two obvious parameters, the central charge $c$ and the level $k$ of the current algebra. 
In subsection \ref{sec:extVirasoro}, we restrict our attention to the case $n=2$; we obtain the OPEs among the currents by requiring the associativity.
Furthermore, we show the uniqueness of the algebra with one parameter, e.g., the level $k$ of the affine symmetry fixed.

\subsection{Rectangular W-algebras}

In this subsection we give a quick overview of the mathematics of rectangular W-algebras. They haven't been studied much, but one reference is \cite{Arakawa:2016fbi}. We use \cite{Kac:2003jh} as reference on quantum Hamiltonian reduction.
Consider a simple Lie algebra $g$, then the chiral algebra of the Wess-Zumino-Novikov-Witten (WZNW) theory of $g$ at level $t$, $g_t$, is the affine vertex algebra of $g$ at level $-t$ (note that normalization of bilinear form between the physics notation that we use and standard notation in vertex algebras differ by a sign and hence we have this sign difference in meaning of level). The notation for the affine vertex algebra is $V_{- t} (g)$. We are interested in the special case $g =$ sl$(L)$.
Consider an embedding $\rho$ of sl$(2)$. This then determines a representation of sl$(2)$ on sl$(L)$ as well as on the standard representation of sl$(L)$.  The decomposition of the standard representation of sl$(L)$ into sl$(2)$-modules can be labelled by a partition of $L$ according to the dimensions of the irreducible summands appearing or equivalently by a Young tableau with $L$ boxes. The Young tableau consists of as many columns as irreducible summands appear in the decomposition of sl$(L)$ and the height of the columns is given by the dimensions of the corresponding irreducible summands. Quantum Hamiltonian reduction then associates a new vertex algebra, a W-algebra, to this data as a certain semi-infinite cohomology. The main point for us is that every lowest-weight vector for the action of sl$(2)$ on sl$(L)$ via $\rho$ gives rise to exactly one generator of the W-algebra of conformal weight $(d+1)/2$ with $d$ the dimension of the corresponding irreducible representation. 

We now further specialize to $L=nM$ and first decompose sl$(nM)$ into a module for $\text{sl}(n) \oplus \text{sl}(M)$ in the obvious way,
\begin{align}
\text{sl}(M) \otimes  \mathbbm{1}_n  \oplus \mathbbm{1}_M \otimes  \text{sl}(n) \oplus  \text{sl}(M) \otimes  \text{sl}(n) \simeq \text{sl}(M n) \, ,
\end{align}
and then consider the embedding of sl$(2)$ in sl$(nM)$ given by the composition of the regular embedding of sl$(2)$ in sl$(n)$ and above embedding of sl$(n)$ in sl$(nM)$. The corresponding Young tableau is then of rectangular type, i.e. it has $M$ columns each of height $n$.
 Especially sl$(nM)$ decomposes as sl$(2)$ representations as
\[
\text{sl}(nM) \cong (M^2-1) \mathbbm{1} \oplus M^2 \mathbbm{3} \oplus \dots \oplus M^2 \mathbbm{2n-1} 
\]
with $\mathbbm{s}$ the $s$-dimensional irreducible representation of sl$(2)$. This rectangular W-algebra is thus of desired type, i.e. it has a sl$(M)$ current algebra and additional $M^2$ fields of spin $2, 3, \dots, n$. The $M^2$ fields of a given spin carry the adjoint plus trivial representation of sl$(M)$. 
Let us denote the W-algebra constructed in this way from the affine vertex algebra of sl$(nM)$ at level $-t$ by $W(-t, n, M)$. The central charge of this $W$-algebra is extracted from equation (2.3) of \cite{Kac:2003jh}. It consists of three summands, the central charge of $V_{-t}(\text{sl}(Mn))$, a dilaton shift in the Cartan direction of the image of $\rho$ of the sl$(2)$ Cartan sub-algebra element $h$ in sl$(nM)$ and various ghost contributions. Note that $\rho(h)$ has norm $Mn(n^2-1)/12$ and so the central charge contribution of the dilaton shift is $  M t n(n^2-1)$.
A pair of ghosts of conformal weight $\lambda, 1-\lambda$ has central charge $1-12(\lambda-1/2)^2$ and
 there are $(n-1)M^2$ ghosts of weight $(1, 0)$, $(n-2)M^2$ ghosts of weight $(2, -1)$ and so on up to $M^2$ ghosts of weight $(n-1, -n+2)$ giving a total contribution to the central charge of  $M^2(n-1)(1-(n-1)^2(n+1))$.
This is then evaluated to
\begin{equation}
\begin{split}
 c_\text{W}(- t, n, M) &= \frac{ t (n^2 M^2-1)}{ t - M n}  + M t n(n^2-1) +M^2(n-1)(1-(n-1)^2(n+1)) \\
  &= \frac{Mn(n^2 M^2-1)}{ t - M n}  +  ( t - Mn) M n(n^2-1) +M^2n (2n^2-1) -1\, .
\label{Wcentern}
\end{split}
\end{equation}
Moreover the level $- \ell$ of the affine vertex algebra of $\text{sl}(M)$ is shifted due to ghost contributions and it is 
\begin{align}
- \ell \, (= k) =  - t  n + M n(n-1) \, .
\label{Wleveln}
\end{align}
This follows since the $M^2n(n-1)$  ghosts carry $n(n-1)$ copies of the adjoint plus trivial representation of sl$(M)$ and each contributing to a level shift by the dual Coxeter number $M$ of sl$(M)$. 
Setting $c = c_\text{W} (- t , n , M)$, we find the relation among $c,k$ as
\begin{align}\label{centraln}
c=	-\frac{\left(k^2-1\right) n^2 M}{k+ n M}+k M-1 \, ,
\end{align}
whose leading behaviour recovers the central charge of the gravity computation  \eqref{c2l2}.

\subsection{Actions for rectangular W-algebras}

We  begin with the simplest case with $n=2$.
We consider the sl$(2 M)$ WZNW model with the level $t$.
The reduction procedure in \cite{Creutzig:2015hla} (see also \cite{Hikida:2007tq,Hikida:2007sz,Creutzig:2011qm})  would lead to the action
\begin{align}
\begin{aligned}
S_t [\varphi , g_1 ,g_2] =& S^\text{WZNW}_{t - M} [g_1] +  S^\text{WZNW}_{t - M} [g_2] 
 \\ &+ \frac{1}{2 \pi} \int d^2 z \left[ \partial \varphi \bar \partial \varphi + \frac{Q_\varphi}{4} \sqrt{g} \mathcal{R} \varphi + \frac{1}{ t}  \text{tr} \left(e^{- 2 b \varphi} g_1^{-1} g_2 \right) \right] \, .
\end{aligned}
\label{actionsl2n}
\end{align}
Here $g_1,g_2$ are the elements of sl$(M)$, and $S^\text{WZNW}_t [g]$ denotes the action of the sl$(M)$ WZNW model with level $t$.
The background charge for $\varphi$ is
\begin{align}
Q_\varphi = - M^2 b - \frac{1}{b} \, , \quad b = \frac{1}{ \sqrt{M (t - 2 M)} } \, .
\end{align}
The central charge of the theory can be computed as
\begin{align}
c = 2 \cdot \frac{(M^2 -1)(t-M)}{t - M - M} + 1 + 6 Q^2 _\varphi 
 = \frac{6 t^2 M - 10 t M^2 - t + 4 M^3}{t-2 M} \, ,
 \label{center2}
\end{align}
which is the same as $c_\text{W} (-t , 2 , M)$ in \eqref{Wcentern}.
Moreover, the diagonal part of the affine symmetries with $g_1$ and $g_2$ survives in the interaction of \eqref{actionsl2n}. Thus the theory admits the symmetry of affine sl$(M)$ with the level 
\begin{align}
\ell \,  =  2 ( t -  M )\, ,
\label{level2}
\end{align}
see \eqref{Wleveln}.

Next, we  study the case with generic $n$.
The realization of rectangular W-algebras as intersection of kernels of screenings of Genra \cite[Prop. 3.7 and Thm. 3.8]{Genra:2016xxc}
suggests that the action is given by
\begin{align}
\begin{aligned}
S_t =& S_t [\varphi_1, \dots, \varphi_{n-1} , g_1 ,\dots, g_n] 
= \sum_{i=1}^n S^\text{WZNW}_{t - M(n-1)} [g_i]  \\
& + \sum_{j=1}^{n-1}\frac{1}{2 \pi} \int d^2 z \left[ \partial \varphi_j \bar \partial \varphi_j + \frac{Q_\varphi}{4} \sqrt{g} \mathcal{R} \varphi_j + \frac{1}{ t}  \text{tr} \left(e^{- 2 b \varphi_j} g_j^{-1} g_{j+1} \right) \right] \, .
\end{aligned}
\label{actionslMn}
\end{align}
The background charge for $\varphi$ is
\begin{align}
Q_\varphi = -\sqrt{\frac{n(n+1)}{6}} \left(M^2 b + \frac{1}{b}\right) \, , \quad b = \frac{1}{ \sqrt{M (t - n M)} } \, .
\end{align}
The central charge of the theory can be computed as
\begin{align}
c = n \cdot \frac{(M^2 -1)(t-M(n-1))}{t - nM} + (n-1) \left( 1 + 6 Q^2 _{\varphi } \right) \,  ,
\end{align}
which simplifies to \eqref{Wcentern}.
The level of the affine sl$(M)$ sub-algebra is
\begin{align}
\ell \,  =  n ( t  - M  (n-1)  )
\label{leveln}
\end{align}
as in \eqref{Wleveln}.

\subsection{The simplest example with $n=2$}
\label{sec:extVirasoro}

As discussed above, the asymptotic symmetry algebra of the higher spin theory includes spin $s$ currents $W^{(s)}$ $(s=2,3,\ldots ,n)$ and the affine su$(M)$ currents $J^a$. 
There are spin $s$ currents $Q^{(s)}_a$ $(s=2,3,\ldots ,n)$ in the adjoint representation of su$(M)$ in addition to $W^{(s)}$ in the trivial one.
We would like to know the OPEs among these currents, which may be determined uniquely by requiring their associativity. In this subsection, we would like to show that this is the case for $n=2$.%
\footnote{At the classical limit, it was shown in \cite{Joung:2017hsi} that the Poisson brackets among generators are rigid by requiring the Jacobi identities with some ansatz.  At $\lambda = 0,1$, free field realizations are possible, and the linear versions of W-algebras with affine su$(M)$ symmetry were obtained in \cite{Bakas:1990xu,Odake:1990rr}.}

The algebra with $n=2$ includes the energy-momentum tensor $T \equiv W^{(2)}$ and the affine su$(M)$ currents $J^a$ with the OPEs
\begin{align}
\begin{aligned}
T (z) T(0) & \sim \frac{c/2}{z^4} + \frac{2 T(0)}{z^2} + \frac{\partial T(0)}{z} \, ,  \\
T (z) J^a (0) & \sim  \frac{J^a (0)}{z^2} + \frac{\partial J^a (0)}{z} \, , \quad
J^a (z) J^b (0)  \sim  \frac{k \delta^{a b}}{z^2} + \frac{i f^{a b }_{~~c} J^c (0)}{z} \, .
\end{aligned}
\label{opeTJ}
\end{align}
Here the structure constant $f^{ab}_{~~c}$ is defined as in \eqref{suMgenerators}.
For a while, we do not assign any relation among the central charge $c$ and the level of su$(M)$ currents $k$.
Along with them, there are also  spin 2 currents $Q^a \equiv Q^{(2)}_a$.
We choose their basis such that the OPEs among $T$ and $J^a$ become
\begin{align}
T (z) Q^a (0) \sim \frac{2 Q^a(0)}{z^2} + \frac{\partial Q^a(0)}{z} \, ,  \quad
J^a (z) Q^b (0)  \sim  \frac{i f^{a b }_{~~c} Q^c (z)}{z} \, .
\label{opeQ}
\end{align}
In the following, we determine the OPEs among $Q^a$ and $Q^b$ by requiring their associativity and show that they are unique.%
\footnote{We utilize the Mathematica package of \cite{Thielemans:1991uw}.}
We first classify  all possible terms generated by the operator products of $Q^a$, and then show that 
the associativity uniquely fixes the coefficients of the terms up to an overall normalization.
In particular, we obtain a relation between the central charge $c$ and the level of su$(M)$ currents as \begin{align}
c = - \frac{ 4  (k^2 - 1)M }{k + 2 M} + k M - 1 \, ,
\label{center2level}
\end{align}
which is the same as the one in \eqref{centraln} with $n=2$.

First we complete the list of operators consisting of $T,J^a,Q^a$ with conformal dimension $h =0,1,2,3$, since the operator product $Q^a (z)Q^b (0)$ is of dimension 4.
The  operator with $h=0$ is the identity, and the operators with  $h=1$ are the su$(M)$ currents $J^a$.
For $h=2$, there are quasi-primary operators such as the energy-momentum tensor $T$, the spin 2 charged currents $Q^a$.
Along with them, there are composite quasi-primary operators
\begin{align}
( J^{(a} J^{b)} ) (z) \equiv \frac{1}{2} \left[ (J^a J^b) (z) +  (J^b J^a) (z) \right] \, .
\end{align}
The normal ordering is defined as
\begin{align}
(A B) (z) = \frac{1}{2 \pi i} \oint \frac{d w}{w - z} A(w) B (z) \, ,
\label{normalorder}
\end{align}
and the brackets $(a_1 , \ldots , a_p)$ and $[a_1 , \ldots , a_p]$ represent the symmetric and anti-symmetric indices, respectively, with the pre-factor $1/( p! )$.
The composite quasi-primary operators with $h=3$ are
\begin{align}
\begin{aligned}
&(J^{(a} J^b J^{c)}) \, , \quad (T J^{a}) - \frac12 \partial^2 J^a \, , \quad 
 (J^a Q^b) - \frac{i}{4} f^{ab}_{~~c} \partial Q^c \, , \\
&\Xi^{ab} = (\partial J^a J^b) - (\partial J^b J^a) - \frac{i}{3}  f^{a b }_{~~ c} \partial^2 J^c \, .
\end{aligned}
\end{align}
There are several ways to make the product $(J^a J^b J^c)$ to be quasi-primary by adding the products of $J^a $.
Here we have chosen one of them with symmetric indices. The differences can be expressed by linear combinations of other quasi-primaries $\Xi^{ab}$.

The operator product $Q^a (z) Q^b (0)$ has two labels $\{a,b\}$, which are symmetric or anti-symmetric depending on the even or odd powers in the $1/z$-expansions. Moreover, we have chosen quasi-primary composite operators such as to include only symmetric or anti-symmetric indices.
Therefore, the coefficients in front of the operators have indices with specific properties, and we would like to classify all of them.
We can construct invariant tensors from the su$(M)$ generators $t^a$ such as
\begin{align}
\text{tr}_M ( [t^a , t^b] t^c ) = i f^{a b c} \, , \quad \text{tr}_M ( \{ t^a , t^b \} t^c ) = d^{a b c} \, , 
\label{structures}
\end{align}
where the indices are totally anti-symmetric for $f^{abc}$ and symmetric for $d^{abc}$.
In terms of these tensors, the products of $t^a$ are written as
\begin{align}
t^a t^b = \frac{\delta^{ab}}{M} \mathbbm{1}_{M} + \frac12 \left( i f^{ab}_{~~c} + d^{ab}_{~~c} \right)t^c \, ,
\end{align}
where we have used \eqref{suMgenerators}.

The invariant tensors are given by products of traces of multiple $t^a$,
which can be expressed in terms of \eqref{suMgenerators} and \eqref{structures}.
The coefficients in front of operators generated should be proportional to them.
Because the su$(M)$ generators are traceless, there is no invariant vector.
For tensors with two indices, we have $\delta^{ab}$ as in \eqref{suMgenerators}.
For  tensors with three indices, we have $f^{abc}$ and $d^{abc}$ defined in \eqref{structures}.
For tensors with label $\{a,b,c,d\}$, we need those with $(ab)(cd)$, $(ab)[cd]$, and $[ab][cd]$.
With the single trace, we have the independent bases 
\begin{align}
\begin{aligned}
&d^{abcd}_{4SS1} \equiv 4 \text{tr}_M (t^{(a} t^{b)} t^{(c} t^{d)}) \, , \qquad
d^{abcd}_{4SS2} \equiv 4 \text{tr}_M (t^{(a} t_{(c} t^{b)} t_{d)}) \, , \\
&d^{abcd}_{4SA} \equiv -  4 i \text{tr}_M (t^{(a} t^{b)} t^{[c} t^{d]}) =  d^{abe} f_{e}^{~cd} \, ,\\
&d^{abcd}_{4AA1} \equiv - 4 \text{tr}_M (t^{[a} t^{b]} t^{[c} t^{d]}) = f^{abe} f_{e}^{~cd} \, , \qquad
d^{abcd}_{4AA2} \equiv 4 \text{tr}_M (t^{[a} t_{[c} t^{b]} t_{d]}) \, ,
\end{aligned}
\end{align}
where we should notice that 
\begin{align}
 \text{tr}_M (t^{(a} t_{[c} t^{b)} t_{d]}) = 0 \, .
\end{align}
There are also tensors consisting of double traces as
\begin{align}
\delta^{ab} \delta^{cd} \, , \quad \delta^{ac} \delta^{bd} \pm \delta^{ad} \delta^{bc} \, .
\end{align}
For tensors with five indices, we need those with $[ab](cde)$.
The two independent bases with single trace are
\begin{align}
\begin{aligned}
&d^{abcde}_{51} \equiv 12 \text{tr}_M ( t^{[a} t^{b]} t^{(c} t^d t^{e)}) \, , \qquad
d^{abcde}_{52} \equiv 12 \text{tr}_M ( t^{[a} t_{(c} t^{b]} t_d t_{e)} ) \, . \\
\end{aligned}
\end{align}
Those with double traces are
\begin{align}
f^{ab}_{~~(c}  \delta_{de)}\, , \quad \delta^{[a}_{~(c} d^{b]}_{~de)}  \, .
\end{align}

With these preparations, we can write down our ansatz for the OPEs;
\begin{align}
&Q^a (z) Q^b (0) \sim \frac{c_1}{c} \delta^{ab} \left( \frac{c/2}{z^4} + \frac{2 a_1 T(0)}{z^2} + \frac{a_1 \partial T (0)}{z} \right) \nonumber \\
& + c_2 f^{ab}_{~~c} \left( \frac{J^c (0)}{z^3} + \frac{1/2 \partial J^c (0)}{z^2} + \frac{a_2 \partial ^2 J^c (0) + a_3 (T J^c) (0)}{z}  \right) \nonumber \\
&+ ( c_{31}\delta^{ab} \delta^{cd} + c_{32} d_{4SS1}^{abcd} + c_{33} d_{4SS2}^{abcd} + c_{34} \delta^{ac} \delta^{bd}) \left( \frac{ (J_{(c} J_{d)} )(0)}{z^2} + \frac{1/2 \partial (J_{(c} J_{d)}) (0)}{z} \right) \nonumber\\
&+ (c_{41} d_{4AA1}^{abcd} + c_{42} d_{4AA2}^{abcd} +c_{43} \delta^{ac}\delta^{bd} ) \frac{\Xi_{cd}}{z} \\
&+ (c_{51}  f^{ab}_{~~c}  \delta_{de}  + c_{52} d_{51}^{abcde} + c_{53} d_{52}^{abcde} +c_{54} \delta^{[a}_{~c} d^{b]}_{~de} ) \frac{(J_{(c}J_d J_{e)})}{z} \nonumber\\
& + c_6 d^{ab}_{~~c} \left( \frac{2 Q^c (0)}{z^2} + \frac{\partial Q^c (0)}{z}  \right)
+ (c_{71} d_{4AA1}^{abcd} + c_{72} d_{4AA2}^{abcd} + c_{73} d_{4SA}^{cdab} + 2 c_{74} \delta^{[a}_{~~c} \delta^{b]}_{~d}) \frac{ (J^c Q^d) (0)}{z}  
\, . \nonumber
\end{align}
Requiring the associativity of OPEs, we find that these coefficients are uniquely fixed up to an overall normalization $\pm \sqrt{c_1}$ as
\begin{align}
&a_{1} = \frac{(2 k+M) \left(k \left(3 k M-2 M^2+1\right)-2 M\right)}{2 (k-1) (k+1) M (3 k+2 M)}\, , \nonumber \\
&a_{2}=\frac{k \left(k \left(3 k M+2 M^2+12\right)+27 M\right)+10 M^2}{6 (k-1) (k+1) M (3 k+2 M)}\, ,\quad
a_3 = - 2 a_2 + \frac13 \, ,\quad
c_2 =  \frac{i c_1}{2k} \, ,\nonumber  \\
&c_{31} = \frac{c_{1} \left(k (4 k+5 M)+2 M^2\right)}{2 (k-1) (k+1) M (2 k+M) (3 k+2 M)}\, ,\quad
c_{32} = \frac{c_{1} k}{4 (k-1) (k+1) (2 k+M)}\, , \nonumber \\
&c_{33}=\frac{c_{1}}{8-8 k^2}\, ,\quad
c_{34}=\frac{c_{1}}{2 k-2 k^3}\, , \quad
c_{41} = \frac{c_{1} (k+M) (k (k M+8)+4 M)}{8 (k-1) k (k+1) M (2 k+M) (3 k+2 M)}\, , \nonumber \\
&c_{42} = 0\, ,\quad
c_{43} = - \frac{c_{1} M}{4 (k-1) k (k+1) (2 k+M)}  \, , \quad 
c_{51}=\frac{i c_{1} (k+M)}{(k-1) k (k+1) M (3 k+2 M)} \,  , \nonumber \\
&c_{52} = \frac{c_{1} k}{12 (k-1) (k+1) (2 k+M) (3 k+2 M)}\, ,\quad
c_{53} = - \frac{c_{1}}{12 (k-1) (k+1) (2 k+M)}\,  , \nonumber \\
&c_{54} = 0\, , \quad
c_{6}= \pm  \frac{i \sqrt{c_{1} k }  (k+M)}{\sqrt{ 2 (k-1) (k+1) (2 k+M) (3 k+2 M)}}\, ,  \quad
 c_{71} = 0 \, ,  \quad
c_{72} = \frac{c_{6}}{k}\, , \nonumber \\
&c_{73} = \frac{i c_{6}}{k+M} \, ,\quad c_{74} = 0 
\end{align}
along with \eqref{center2level}. 
We have checked this for $M=2,3,4,5$ and claim that the above expressions also hold for $M > 5$.

All bases used above are independent for $M=4,5$, but this is not the case for $M=2,3$.
Therefore, we should be careful to claim the uniqueness of the algebra. 
For $M=3$, we have
\begin{align}
&   d_{4SS2}^{abcd} =  2 \delta^{ab} \delta^{cd} - 2 d_{4SS1}^{abcd} +   2  ( \delta^{ac} \delta^{bd} + \delta^{ad} \delta^{bc})  \, ,
& d_{51}^{abcde} = 3 i f^{ab}_{~~(c}  \delta_{de)}  \, .
\end{align}
Using these relations, we can set, say, $c_{33}$ and $c_{51}$ arbitrarily by changing other parameters.
There is no other freedom to tune parameters and the associativity uniquely fixes them as in the above expressions.
For $M=2$, we have more relations.  Using
\begin{align}
d^{abc} = 0 \, ,\quad d_{4 AA 2}^{abcd} = 0 \, ,  \quad d_{4SA}^{abcd} = 0 \, , 
\end{align}
we can use arbitrary $c_6 ,c_{53} , c_{72}, c_{42} , c_{73}$.
There are more relations as
\begin{align}
\begin{aligned}
&d_{51}^{abcde} = - d_{52}^{abcde} = 3 i f^{ab}_{~~(c} \delta_{de)} \, , \quad d^{abcd}_{4AA1} = 2 (\delta^{ac} \delta^{bd} - \delta^{ad} \delta^{bc}) \, , \\ 
&d^{abcd}_{4SS1} = 2 \delta^{ab} \delta^{cd} \, , \quad
d^{abcd}_{4SS2} = - 2  \delta^{ab} \delta^{cd} + 2 (\delta^{ac} \delta^{bd} + \delta^{ad} \delta^{bc}) \, . 
\end{aligned}
\end{align}
With them, we can set, say, $c_{52},c_{54},c_{41},c_{71},c_{32},c_{33}$ arbitrarily by changing other parameters as in the above expressions. Others are fixed by the associativity of OPEs as for other $M$.

\section{Symmetry of Grassmannian-like coset}
\label{sec:Grassmannian}

In the previous section, we have examined the asymptotic symmetry of the higher spin gravity with $\lambda =n$ beyond the classical limit.
The holographic duality of \cite{Creutzig:2013tja} suggests that the symmetry can be identified with that of the coset \eqref{Grassmann} with $\lambda = n$. In the next subsection, we confirm this proposal by comparing the central charge $c$ of the model and the level $k$ of the su$(M)$ current. In subsection \ref{sec:Qa}, we explicitly construct the spin 2 currents in terms of the coset model and reproduce the OPEs among them for $n=2$.
In subsection \ref{sec:decomposition}, we provide an interpretation of the map of parameters by decomposing the coset algebra.

\subsection{Map of parameters}
\label{sec: rectangular}

The symmetry algebra of the coset \eqref{Grassmann} includes the energy-momentum tensor, which can be obtained from the standard coset construction. The central charge is computed as
\begin{align}
\begin{aligned}
c_\text{coset}(k, N, M)
 &= \frac{((N+M)^2 -1)k}{k + N + M} - \frac{(N^2 -1)k}{k + N } - 1  \\
&= \frac{k M (k (M+2 N)+N (M+N)+1)}{(k+N) (k+M+N)}-1 \, .
\end{aligned}
\label{ccoset}
\end{align}
The coset also has the symmetry generated by su$(M)$ currents with level $k$. They come from su$(M)$ within su$(N+M)$ in the numerator of \eqref{Grassmann}, which are regular with respect to the action of the currents in the denominator. 
In terms of 't Hooft parameter in \eqref{tHooft}, the central charge \eqref{ccoset} is written as
\begin{align}
c=	-\frac{\left(k^2-1\right) \lambda^2 M}{k+\lambda M}+k M-1 \, ,
\label{centertHooft0}
\end{align}
which reduces to \eqref{c2l} at the 't Hooft limit with $k$ large but $\lambda,M$ fixed.
Therefore, we find the agreement of $c,k$ at the 't Hooft limit as shown in \cite{Creutzig:2013tja} for the $\mathcal{N}=2$ supersymmetric case.

We would like to extend the comparison beyond the 't Hooft limit with setting $\lambda =n$ as before.
We require that the level $k$ of the affine su$(M)$ and the central charge $c_\text{coset}(k, N, M)$ are the same as \eqref{Wleveln} and \eqref{Wcentern} obtained from the quantum Hamiltonian reduction of $\text{sl}(M n)$.
That is, for fixed $n, M, N$, we look for the solutions for $k$ such that
\begin{equation}
k = - t n+Mn(n-1) \, ,\quad 
 c_\text{coset}(k, N, M)  = c_\text{W} (- t, n, M) \, .
\end{equation}
Removing $t$, the above condition can be written as 
\begin{align}
 \frac{\left(k^2-1\right) M (k (n-1)+n N) (k (n+1)+n (M+N))}{(k+N) (k+M n) (k+M+N)}=0 \, .
\end{align}
For $M \neq 0$, the solutions are 
\begin{align}
 k =1 ,-1 , - \frac{n N}{n - 1} ,  - \frac{n (N + M)}{n+1}  \, .
\end{align}
The first two solutions correspond to the free fermion and $\beta\gamma$-ghost  realization of the coset \eqref{Grassmann} as explained in, e.g., \cite{DiFrancesco:1997nk}, where the fields transform as the fundamental and anti-fundamental representations of su$(M)$ and su$(N)$, respectively. In these cases one expects that the simple rectangular $W$-algebra is just the current algebra or a conformal extension of the current algebra, see \cite[Remark 5.3]{Creutzig:2016ehb}, \cite[Section 5.1.1]{Creutzig:2017qyf} and \cite{2018arXiv180509771A} for the example $k=-1$ especially at $n=M=2$.

The non-trivial solutions are thus the latter two.
The third solution can be written as $n = k/(N+k)$, which implies that $\lambda = n$ from \eqref{tHooft}.
We may compare the expressions of central charge in \eqref{centraln} and \eqref{centertHooft0}.
This case may be explained from the AdS/CFT correspondence at the 't Hooft limit as a kind of analytic continuation.
The fourth solution may be expressed as $n = - k/(k + N + M)$, which motivates us to define another 't Hooft parameter \eqref{tHooftp}.
In fact, the central charge of the coset \eqref{Grassmann} in terms of $\lambda'$ is given by \eqref{centertHooft0} with $\lambda$ replaced by $\lambda '$.
This implies that there is a duality relation as in \cite{Gaberdiel:2012ku}. 
We will make some comments on the duality relation soon but let us first comment on this matching of central charges and levels.

\subsection{Singular vectors for boundary admissible level theories}

This very nice matching of central charges and levels of course leads us to conjecture that the coset at these special levels is precisely isomorphic to the corresponding rectangular W-algebra. Proving such a conjecture is rather difficult. On the other hand in order for the conjecture to be true the coset has to have singular vectors at conformal weight $n+1$, which of course implies that the affine vertex algebra of su$(N+M)$ with this $ k$ has singular vectors at conformal weight $n+1$. Unfortunately the level $- n N / ( n - 1)$ is only an admissible level if $n\geq 2(M+N)/M$ and it is rather difficult to study. On the other hand the level $ -  n (N + M) /( n+1 )$ is always a boundary admissible level of su$(N+M)$ \cite{KWbdy}.

Since admissible level WZW theories are not as well-known in the CFT community as the positive integer theories we decided to start with a small overview of features and known results. The point is that admissible level theories have so-called ordinary modules that form a tensor subcategory of modules that is finite and semi-simple, i.e. very similar to the positive integer case. But there are many more modules that are  not always completely reducible
and thus lead to logarithmic theories, see \cite{Creutzig:2013hma} for an introduction and \cite{MR1359963,MR3093193,MR3333645} for the example of sl(2) at admissible level. 
We are here only interested in ordinary modules. Firstly, a level $k$ is called principal admissible if it is a rational number  satisfying $k+h^\vee \geq  h^\vee /u$ and $u$ is both coprime to the dual Coxeter number $h^\vee$ of the Lie algebra $\mathfrak{g}$ as well as to its lacety $r^\vee$ \cite{MR1026952}. The category of ordinary modules consists of simple highest-weight modules at level $k$ of those weights that are also weights of the WZW theory of $\mathfrak g$ at level $\ell = u-h^\vee$ \cite[Main Theorem]{arakawa2016}. Characters of modules can be meromorphically continued to components of vector-valued meromorphic Jacobi forms, i.e. they have some nice modularity properties \cite{MR949675}. However these characters do not form a closed representation of the modular group and one has to include characters of more modules. Nonetheless, the tensor category of these ordinary modules is nice in the sense that they form a braided fusion category and a weak version of Verlinde's formula holds \cite{MR3845289,2018arXiv180700415C}.

Using the character formulae of Kac and Wakimoto one can thus possibly detect singular vectors of WZW theories at admissible level. This works particularly well if we are at the boundary of admissibility, i.e. $k+h^\vee= h^\vee / u $. Fortunately, we are in that situation and
by Remark 1 of \cite{KWbdy} the character of su$(N+M)$ at level $ k= -  n (N + M) /( n+1)$ is just
\begin{equation}
\text{ch}[\text{su}(N+M)_{- \frac{n (N + M)}{n+1}}] = q^{n d}\frac{\Pi(u, \tau)}{\Pi(u, (n+1)\tau)}, \qquad d= (N+M)^2-1
\end{equation}
with $\Pi(u, \tau)$ the vacuum character of the universal affine vertex algebra of su$(N+M)$, that is
\begin{equation}
\Pi(u, \tau) = \frac{1}{\prod\limits_{n=1}^\infty (1-q^n)^{N+M-1} \prod\limits_{\alpha \in \Delta_+} (1-z^\alpha q^n)(1-z^{-\alpha} q^{n-1})}
\end{equation}
where $\Delta_+$ denotes the set of positive roots and $z^\alpha$ is short-hand for $e^{2\pi i \alpha(u)}$ and $u$ in the Cartan sub-algebra of su$(N+M)$. 
We thus see that the character $\text{ch}[\text{su}(N+M)_{- \frac{n (N + M)}{n+1}}]$ agrees with the character of the universal affine vertex algebra up to 
conformal weight $n$ but at conformal weight $n+1$ they differ by the character of the adjoint representation of the finite dimensional Lie algebra su$(N+M)$. 
There  must thus be a singular vector at conformal weight $n+1$ in this representation and so especially there are $M^2$ null fields of conformal weight $n+1$ that decouple. 
In principle it would be nice to compute the precise coset character and this would amount to computing the Fourier coefficient of a negative index meromorphic Jacobi form in $N$-variables. Technology is only available for the one variable case \cite{BCR14, BRZ16}, but the higher rank case is work in progress so that we hopefully can return to a more detailed character analysis in the future. 

\subsection{Charged spin 2 currents}
\label{sec:Qa}

In \cite{Creutzig:2013tja} (see also \cite{Candu:2013fta}), the match of partition function was shown for the holographic duality at the 't Hooft limit. This, in particular, implies that the spin contents of the symmetry generators agree with each other.
In this subsection, we obtain the exact expression of spin 2 currents $K^a$ with finite $c$ in terms of the coset model \eqref{Grassmann}.
We require the OPEs as in \eqref{opeQ}, which shall uniquely determine the form of the spin 2 currents up to an overall factor.

In order to obtain the explicit form of generators, we set up our notation of su$(N+M)$.
We decompose su$(N+M)$ as
\begin{align}
	\text{su}(N+M) = \text{su} (N) \oplus \text{su} (M) \oplus \text{u} (1) 
	\oplus (N , \bar{M}) \oplus (\bar N , M) \, ,
\end{align}
where $L$ and $\bar{L}$ denote the fundamental and anti-fundamental representations of su$(L)$.
We use the generators $t^A = (t^\alpha , t^a , t^{\text{u}(1)}, t^{(\rho \bar \imath)} , t^{(\bar \rho i)})$, respectively.
We express the generators in terms of $(N+M) \times (N+M)$ matrices and use the trace for the matrices as $\text{tr} = \text{tr}_{N+M}$. We use the normalization of generators such that the metric $g^{AB} = \text{tr} (t^A t^B)$ becomes
\begin{align}
	\text{tr} (t^\alpha t^\beta) = \delta^{\alpha  \beta} \, , \quad
	\text{tr} (t^a t^b) = \delta^{a b} \, , \quad
	\text{tr} (t^{\text{u}(1)}t^{\text{u}(1)}) = 1 \, , \quad
	\text{tr} (t^{(\rho \bar \imath)}t^{(\bar \rho i)}) = \delta^{\rho \bar \rho} \delta^{i \bar \imath } \, .
\end{align}
We also need the invariant tensors
\begin{align}
 \text{tr} ([t^A , t^B] t^C) = 	i f^{ABC}  \, , \quad 
 \text{tr} (\{ t^A , t^B\} t^C) =	d^{ABC} 
\end{align}
as in \eqref{structures}.
Several explicit expressions are (see, e.g., appendix B of \cite{Creutzig:2014ula})
\begin{align}
		&i f^{(\rho \bar \imath) (\bar \sigma j) \text{u}(1)} = \sqrt{\frac{M+N}{MN}} \delta^{j \bar \imath } \delta^{\rho \bar \sigma} \, , \quad
		d^{(\rho \bar \imath) (\bar \sigma j) \text{u}(1)} = \frac{M-N}{\sqrt{MN(N+M)}} \delta^{j \bar \imath } \delta^{\rho \bar \sigma} \, , \\
		&i f^{(\rho \bar \imath) (\bar \sigma j) \alpha} = d^{(\rho \bar \imath) (\bar \sigma j) \alpha} = \delta^{\rho \bar \rho} \delta^{\sigma \bar \sigma} (t^\alpha)_{\sigma \bar \rho}  \delta^{j \bar \imath }  \, , \quad
		i f^{(\rho \bar \imath) (\bar \sigma j) a} = - d^{(\rho \bar \imath) (\bar \sigma j) a} = - \delta^{\rho \bar \sigma} (t^a)_{i \bar \jmath}  \delta^{i \bar \imath } \delta^{j \bar \jmath }  \, . \nonumber
\end{align}
Important properties of the invariant tensors may be found in appendix B of \cite{Bais:1987dc}.

With the preparation, we can write down low spin generators explicitly.
The spin one generators are the su$(M)_k$ currents given by $J^a$.
The energy-momentum tensor can be constructed as
\begin{align}
	T = \frac{g_{AB}}{2(k + N + M)} (J^A J^B) - \frac{1}{2 (k+ N)} (J^\alpha J^\alpha) - \frac{1}{2 kNM(N+M)} (\tilde J^{\text{u}(1)} \tilde J^{\text{u}(1)}) \, ,
	\label{emt}
\end{align}
where $\tilde J^{\text{u}(1)} = \sqrt{NM (N + M)} J^{\text{u}(1)}$.
Here the normal ordering is defined as in  \eqref{normalorder}.
We look for symmetry generators, which are regular with respect to the action of the currents in the denominator of \eqref{Grassmann}. 
The spin 2 generators $K^a$ in the adjoint representation of su$(M)$ should be given by linear combinations of
\begin{align}
 ( J^b J^c ) \, , \quad  (J^{(\rho \bar \imath)} J^{(\bar \rho j)} ) \delta_{\rho \bar \rho} \, , \quad   (J^{(\bar \rho j)} J^{(\rho \bar \imath)}) \delta_{\rho \bar \rho} \, , \quad   (J^a J^{\text{u}(1)}) \, , \quad \partial J^a \, .
\end{align}
We would like to have linear combinations which are consistent with the OPEs in \eqref{opeQ}.
It is useful to use the formula
\begin{align}
	\begin{aligned}
		J^C (z) K^{AB} (0) \sim &\frac{i k f^{CA}_{~~~D} g^{DB}  }{z^3} + 
		\frac{k g^{CA} J^B (0) +k g^{CB} J^A (0) - f^{CA}_{~~~D} f^{DB}_{~~~F} J^F (0)}{z^2}
		\\ & + \frac{i f^{CA}_{~~~D} K^{DB}  (0) + i f^{CB}_{~~~D}  K^{AD}(0)}{z} \, ,
	\end{aligned}
\end{align}
which can be obtained from
\begin{align}
	J^A (z) J^B (0) \sim \frac{k g^{AB}}{z^2} + \frac{i f^{AB}_{~~~C} J^C (0)}{z } \, , \quad 
	K^{AB} (z) = (J^A J^B) (z) \, .
\end{align}
We then obtain the expression of spin 2 currents as
\begin{align}
\begin{aligned}
	K^a =&   [ (J^{(\rho \bar \imath)} J^{(\bar \rho j)})  +  (J^{(\bar \rho j)} J^{(\rho \bar \imath)}) ]\delta_{\rho \bar \rho} t^a_{j \bar \imath} \\
	 & \qquad - \frac{N}{M + 2 k} d^{a b c} (J^b J^c) + \frac{2}{k} \sqrt{\frac{N (N+M)}{M}} ( J^a J^{\text{u}(1)}) 
\end{aligned}
\end{align}
up to an overall normalization. 
In a similar manner, we can construct spin 3 currents in the coset language, see appendix \ref{sec:spin3}.

With the explicit expression of $K^a$ in terms of su$(N+M)$ currents, we can compute the OPEs among them. In general, the operator products produce spin 3 currents, and the new operators would generate higher spin currents as well.
However, at $\lambda =2$, the operator products of spin 2 currents $Q^a$ in the rectangular W-algebra do not generate new spin 3 currents, and in fact we computed that the algebra is uniquely fixed by requiring the associativity of OPEs.
Therefore, once we accept the decoupling of the spin 3 currents, then the operator products of $K^a$ and those of $Q^a$ should be identical. We have computed the operator products of $K^a$ with explicit values of $N,M$%
\footnote{We have computed with $N=2,\ldots,6$ and $M=2, \ldots , 5$ except for $M=N$ where the central charge diverges.}
 and deduced the relation as
\begin{align}
K^a (z) K^b (0) \sim Q^a (z) Q^b (0) + \frac{i f^{a b }_{~~c}P^c (0)}{z}
\end{align}
at $\lambda = 2$. Here $P^a$ are given by
\begin{align}
P^a = 4 ( J^\alpha ( J^{\rho \bar \imath} J^{\bar \sigma j} )) t^\alpha_{\rho \bar \sigma} t^a_{j \bar \imath} - \frac{2}{ N} ( J^\alpha ( J^{\alpha} J^{a} )) \, . 
\label{spin3critical}
\end{align}
We can check that they satisfy
\begin{align}
T (z) P^a (0) \sim \frac{3 P^a(0)}{z^2} + \frac{\partial P^a(0)}{z} \, ,  \quad
J^a (z) P^b (0)  \sim  \frac{i f^{a b }_{~~c} P^c (z)}{z} \, ,
\label{opeP}
\end{align}
which means that they can be interpreted as the charged spin 3 currents.
See appendix \ref{sec:spin3} for the expressions with generic $k$.
A crucial point here is that the operator products of $P^a$ are of the form
\begin{align}
	P^a (z) P^b (0) \sim \frac{0 \cdot \delta^{ab}  }{z^6} + \mathcal{O} (z^{-5}) \, .
\end{align}
That is, the states corresponding to the spin 3 currents have zero norm, and hence they can be regarded as null vectors. Therefore, we can decouple $P^a$ by setting $P^a = 0$.
In this way, we can confirm the equality of the symmetries from the affine sl$(2 M)$ and the coset  \eqref{Grassmann} with $\lambda = 2$.

\subsection{Decomposition of rectangular W-algebra}
\label{sec:decomposition}

In subsection \ref{sec: rectangular}, we argued that the W$_n$-algebra with su$(M)$ symmetry can be realized as the symmetry of the coset \eqref{Grassmann} if $ k/(k+N) = n $ or $ - k /(k + N +M) = n$. 
In this subsection, we would like to examine the meaning of the map of parameters more closely.
In order to do so, it is convenient to decompose the coset algebra as
\begin{align}
\begin{aligned}
\frac{\text{su}(N+M)_k}{\text{su}(N)_k \oplus \text{u}(1)} \supset
 \frac{\text{su}(N+M)_k}{\text{su}(N+M-1)_k \oplus \text{u}(1)} 
\oplus \cdots \oplus  \frac{\text{su}(N+1)_k}{\text{su}(N)_k \oplus \text{u}(1)}  \\ \oplus ( M-1 )  \text{u}(1)\, .
\end{aligned}
\label{cosetdec0}
\end{align}
The decomposition can be nicely explained in terms of brane junctions \cite{Gaiotto:2017euk,Creutzig:2017uxh,Prochazka:2017qum,Prochazka:2018tlo,Harada:2018bkb}.
We would like to examine the properties of the coset \eqref{Grassmann} in terms of the sum of the cosets.

Each coset in \eqref{cosetdec0} can be interpreted in terms of  W$_\infty[\lambda]$ with a specific value of $\lambda$.
Here we define W$_\infty[\lambda]$ as the quantum Hamiltonian reduction of hs$[\lambda]$ with the principal embedding of sl(2). 
Because of the truncation of hs$[\lambda]$, 
the algebra W$_\infty [\lambda]$ can be reduced to W$_L$ at $\lambda = L$. 
The W$_L$-algebra can be realized as the symmetry of the coset (finally proven in \cite[Main Theorem 1 and 2]{Arakawa:2018iyk})
\begin{align}
	\frac{\text{su}(L)_K \oplus \text{su}(L)_1}{\text{su}(L)_{K+1}}
	\label{basiccoset}
\end{align}
with the central charge
\begin{align}
	c_{L,K} = (L-1) \left(1-\frac{L (L+1)}{(K+L) (K+L+1)}\right) \, .
\end{align}
In \cite{Gaberdiel:2012ku} (see also \cite{Prochazka:2014gqa}), a triality relation of W$_\infty [\lambda]$ was found, and the algebra  was suggested to be the same with the three choices of $\lambda$ as
\begin{align}
\lambda_1 = \frac{L}{L + K} \, , \quad \lambda_2 = - \frac{L}{L + K + 1} \, ,	 \quad  \lambda_3 = L 
	\label{lambdas}
\end{align} 
and with fixed $c = c_{L,K}$. 
There is an isomorphism between the universal enveloping algebra of W$_\infty[\lambda]$ and the affine Yangian of gl$(1)$ in \cite{Tsymbaliuk} as explained in \cite{Prochazka:2015deb,Gaberdiel:2017dbk}. In the Yangian description, there are types of representation expressed by plane partitions. We need three axes, e.g., $x_1,x_2,x_3$ to express the plane partitions, and the invariance under the rotation of axes corresponds to the triality relation.

It has been known for a long time \cite{Altschuler:1989nm,Walton:1988bs}, see \cite[Theorem 13.1]{Arakawa:2018iyk} for a proof, that the coset 
\begin{align}
\frac{\text{su}(K+1)_ L }{\text{su}(K)_L \oplus \text{u}(1)}
\end{align}
is level-rank dual to the coset \eqref{basiccoset}. 
Therefore, the decomposition of \eqref{cosetdec0} implies that
\begin{align}
\begin{aligned}
\frac{\text{su}(N+M)_k}{\text{su}(N)_k \oplus \text{u}(1)}
\supset  \text{W}_\infty [\mu^{(M-1)}] \oplus \cdots \oplus  \text{W}_\infty [\mu^{(0)}] \oplus (M-1) \text{u}(1) \, ,
\end{aligned}
\label{cosetdec}
\end{align}
where the parameters $\mu^{(i)}$ are one of three possibilities
\begin{align}
\mu^{(i)}_1 = \frac{k}{k + N + i} \, , \quad
\mu^{(i)}_2 = - \frac{k}{k + N + i + 1} \, , \quad 
\mu^{(i)}_3 = k \, .
\end{align}
Here the each W-algebra is labeled with the central charge $C_i = c_{k , N + i}$.
We can check that the central charge of the coset \eqref{Grassmann} can be reproduced from the right hand side of \eqref{cosetdec} as
\begin{align}
c_\text{coset} (k,N,M) = \sum_{i=0}^{M-1} C_i + M-1 \, .
\end{align}
In appendix \ref{sec:webofwalg}, we explicitly decompose the symmetry algebra with $\lambda =2$ and $M=2$ as in \eqref{cosetdec} using the results obtained in subsection \ref{sec:extVirasoro}.

In order to express the whole coset algebra in terms of the sum of component W-algebras, we need to consider operators connecting two of the components. The symmetry algebra should be generated by currents with integer conformal dimensions. 
In the present case, the conformal dimensions of operators connecting two W-algebras can be integer with the help of the relation
\begin{align}
 \mu_2^{(i)} + \mu_1^{(i+1)} = 0 \, ,
 \label{lambdarel}
\end{align}
see discussions in subsection 4.4.1 of \cite{Prochazka:2017qum}. 
In the Yangian description, we need to connect plane partitions with coordinates $(x_1^{(i)},x_2^{(i)},x_3^{(i)})$ \cite{Prochazka:2017qum,Gaberdiel:2017hcn,Gaberdiel:2018nbs}.
The relation \eqref{lambdas} implies that the $x_2^{(i)}$-direction should be connected with the  $x_1^{(i+1)}$-direction for all $i=0,1,\ldots , M-1$.
Therefore, among the product of the rotational symmetries of axes, only $\mathbb{Z}_2$-symmetry reversing the connected direction is remained. 
This exchanges, in particular, $\mu_1^{(0)}$ and $\mu_2^{(M-1)}$, which explain the duality of the coset models \eqref{Grassmann} with $\mu_1^{(0)} = \mu$  and $\mu_2^{(M-1)} = \mu$ ($\mu$ is a number) but with a fixed $k$.

At the 't Hooft limit, all the component W-algebras become the same as W$_\infty [\lambda]$ with the central charge $c=(1 - \lambda^2) k$. The same result can be obtained from the higher spin theory by following the analysis in section \ref{sec:Vasiliev}. In this case, we restrict ourselves to the sub-sectors generated by $t^H \otimes V^s_m$ $(H=1,2,\ldots,M)$, where $t^H$ are generators of the Cartan sub-algebra of gl$(M)$.
At $\lambda =n$  with $n = 2,3, \ldots$, we have $\mu^{(0)}_1 = n$, thus the component W$_\infty [\mu^{(0)}]$ can be truncated to be a W$_n$-algebra.
A similar statement holds also for $\lambda ' = n$.
This fact should be related to the truncation of the higher spin currents as argued in subsection \ref{sec: rectangular}.

A unitary model can be realized when $k=n$ with an integer $n$, where all W$_\infty [\mu^{(i)}]$ can be truncated to W$_{k}$-algebra.
However, this is not a good choice in order to realize the W-algebra with su$(M)$ symmetry, since the level $k$ of su$(M)$ becomes the same as the upper bound for the spin of currents. 
For $M=1$, the restriction from the level of su$(M)$ disappears, and we can set $k=n$ as
\begin{align}
\frac{\text{su}(N+1)_n}{\text{su}(N)_n \oplus \text{u}(1)} \simeq \frac{\text{su}(n)_N \oplus \text{su}(n)_1}{\text{su}(n)_{N+1}} \, .
\end{align}
However, we may consider an analytic continuation of $N$ in order to relate with the symmetry of sl$(n)$ Chern-Simons theory \cite{Castro:2011iw,Gaberdiel:2012ku,Perlmutter:2012ds}.

\section{Conclusion}
\label{sec:conclusion}

In this paper, we examined the asymptotic symmetry of the 3d Prokushkin-Vasiliev theory with $M \times M$ matrix valued fields. The gauge algebra is given by hs$_M [\lambda]$ defined in \eqref{hsMlambda}, and it can be truncated to be sl$(M n)$ at $\lambda = n$ as in \eqref{slMn}. At $\lambda =n$, the symmetry algebra is obtained by the Hamiltonian reduction of sl$(M n)$ with the sl(2) embedding corresponding to the partition \eqref{partition}.
The W-algebra includes the affine su$(M)$ Lie algebra as a sub-algebra. We computed  the central charge $c$ of the W-algebra and the level $k$ of the affine su$(M)$
as in \eqref{Wcentern} and \eqref{Wleveln}.  We first obtained the expressions at the classical limit from the Chern-Simons description of higher spin theory and then found the exact results from the quantum Hamiltonian reduction of sl$(M n)$.
For $n=2$, we obtained the OPEs of the symmetry generators by requiring their associativity.

Based on the holographic duality in \cite{Creutzig:2013tja}, we claim that the symmetry algebra is the same as the one of the Grassmannian-like coset in \eqref{Grassmann} if the 't Hooft parameter in \eqref{tHooft} or in \eqref{tHooftp} is set as $\lambda = n$ or $\lambda ' = n$. We obtained the map of parameters from the comparison of  the central charge $c$ of the algebra and the level $k$ of the affine su$(M)$. We constructed low spin currents explicitly in terms of the coset model and reproduce OPEs for $n=2$.
In order to interpret the meaning of the 't Hooft parameters, we decomposed the W-algebra as in \eqref{cosetdec}. 
We can explain the duality of the coset model \eqref{Grassmann} by combining the triality relation of \cite{Gaberdiel:2012ku} and the Yangian description of \cite{Prochazka:2015deb,Gaberdiel:2017dbk} for the component W-algebras.

We need further study in order to confirm our claim on the match of the symmetry algebra with generic $n >2$.
We may be able to obtain the OPEs of higher spin by requiring their associativity, and this would lead to the relation of $c,k$ in \eqref{centraln} with generic $n$. If we work with generic $\lambda$, the duality relation would be derived from the invariance of the structure constants as in \cite{Gaberdiel:2012ku,Prochazka:2014gqa,Candu:2012tr}.
The symmetry algebra could be reconstructed from the coset model as in subsection \ref{sec:Qa}.
For the purpose, we need the explicit form of higher spin currents and the OPEs among these currents. 
Since the explicit computations are straightforward but quite tedious, it would be better if there is an abstract way to prove our proposal.
In subsection \ref{sec:decomposition}, we briefly comment on the Yangian description of the W-algebra with su$(M)$ symmetry, but we have not worked out any details there. It is important to develop the description in order to understand the properties of the W-algebra with deformable parameters. In particular, we would like to understand more the truncation of spectrum at $\lambda = n$. We also want to construct conical defect geometry in the sl$(Mn)$ Chern-Simons theory and compare them to some states in the W-algebra as in \cite{Castro:2011iw,Gaberdiel:2012ku,Perlmutter:2012ds,Hikida:2012eu}.
We hope to report on some developments in near future.

We have examined the matrix extension of higher spin gravity in order to see the relation to string theory.
In order to proceed furthermore, it would be useful to introduce an extended supersymmetry.
In our previous works \cite{Creutzig:2014ula,Hikida:2015nfa}, we introduced the $\mathcal{N}=3$ supersymmetry by dealing with a critical level of Grassmannian model.
For the extended supersymmetry, we embedded the Clifford algebra into the  matrix algebra of higher spin theory as in \cite{Prokushkin:1998bq,Henneaux:2012ny}. Thus, it is natural to expect that the analysis in this paper can be applied to the case with extended supersymmetry in a rather straightforward manner.
The $\mathcal{N}=4$ supersymmetry was introduced in \cite{Gaberdiel:2013vva,Gaberdiel:2014cha,Gaberdiel:2015mra}, 
and the relation to the symmetric orbifold was discussed. The symmetry of the symmetric orbifold was named as "higher spin square," and it is interesting to see how the algebra is related to the one examined in this paper.

\subsection*{Acknowledgements}

We are grateful to Boris Feigin, Andrew Linshaw,  Sanefumi Moriyama, Satoru Odake and Takahiro Uetoko for useful discussions.
The work of YH is supported by JSPS KAKENHI Grant Number 16H02182.
The work of TC is supported by NSERC grant number RES0019997.
\appendix

\section{Spin 3 currents from the coset}
\label{sec:spin3}

In this appendix, we obtain spin 3 currents from the coset model \eqref{Grassmann}.
The currents should be constructed from composite operators of the currents in the numerator, which are regular with the currents $J^\alpha , J^{\text{u}(1)}$ in the denominator.
We first study the charged ones and then examine the singlet one with respect to the su$(M)$ action.

For the charged currents, we choose the  composite operators transforming in the trivial and adjoint representations of su$(N)$ and su$(M)$, respectively, and use the ansatz
\begin{align}
P^a &= a_1 (J^\alpha (J^{\rho \bar \imath} J^{\bar \sigma j})) t^\alpha_{\rho  \bar \sigma} t^a_{j \bar \imath}
+ a_2 (J^\alpha (J^\alpha J^a)) + a_3 (J^b( J^b J^a)) + a_4 (J^a (J^{\text{u}(1)} J^{\text{u}(1)}) ) \nonumber  \\
&+  (a_5 d^{abc} + a_6  i f^{abc}) \left[ (J^c (J^{\rho \bar \imath} J^{\bar \rho j}) )+ (J^c (J^{\bar \rho j} J^{\rho \bar \imath} ) )\right] \delta_{\rho \bar \rho}  t^b_{j \bar \imath} \nonumber \\
&+ a_7 \left[ (J^{\text{u}(1)} (J^{\rho \bar \imath} J^{\bar \rho j})) + (J^{\text{u}(1)} (J^{\bar \rho j} J^{\rho \bar \imath} ) ) \right] \delta_{\rho \bar \rho}  t^a_{j \bar \imath} 
+ a_8 \left[ (J^{a} (J^{\rho \bar \imath} J^{\bar \rho i})) + (J^{a} (J^{\bar \rho i} J^{\rho \bar \imath} ) ) \right] \delta_{\rho \bar \rho} \delta_{i \bar \imath} \nonumber \\
&+ a_9 d^{abc} (J^b (J^c J^{\text{u}(1)})) +( a_{10} d^{abc}+ a_{11} i f^{abc}) ((\partial J^b) J^c )  \\
&+ \left[ a_{12} ((\partial J^{\rho \bar \imath}) J^{\bar \rho j}) + a_{13}((\partial  J^{\bar \rho j}) J^{\rho \bar \imath}) \right] \delta_{\rho \bar \rho} t^a_{j \bar \imath} 
+a_{14} (J^a (\partial J^{\text{u}(1)})) + a_{15}  ((\partial J^a )J^{\text{u}(1)}) \nonumber \\
&+ a_{16} \partial ^2 J^a + a_{17} 6 \text{tr}_M  (t^a t^{(b} t^c t^{d)}) (J^b (J^c J^d)) \, . \nonumber
\end{align}
Along with the condition with $J^\alpha , J^{\text{u}(1)}$, we require
for these currents to satisfy the OPEs \eqref{opeP}.
Then we can fix the coefficients as
\begin{align}
&a_{2} =  \frac{a_1}{k} \, , \quad a_{3} =  \frac{a_1 N (k+2 N)}{k (k+M) (3 k+2 M)} \, , \quad a_{4} =  \frac{a_1 (k+2 N) (M+N)}{k^2 M} \, ,  \nonumber \\
& a_{5} =  -\frac{a_1 (k+2 N)}{4 (k+M)} \, , \quad  a_{6} =  0 \, , \quad a_{7} = \frac{a_1 (k+2 N) }{2 k } \sqrt{\frac{M+N}{M N}} \, , \quad a_{8} = -\frac{a_1 (k+2 N)}{2 k M} \, ,   \nonumber\\
&  a_{9} =  -\frac{a_1 (k+2 N) }{2 k (k+M)} \sqrt{\frac{N (M + N)}{ M}} \, , \quad  a_{10} =  0  \, , \quad a_{11} =  \frac{a_1 \left(k^2-8\right) N (k+2 N)}{4 k (k+M) (3 k+2 M)}  \, , \\
 &  a_{12} =  -\frac{a_1 (k+2 N)}{2} \, , \quad a_{13} =  \frac{ a_1 (k+2 N)}{2} \, , \quad a_{14} =  0 \, , \quad a_{15} =  0 \, ,  \nonumber\\ 
&a_{16} =  -\frac{a_1 N \left(6 k^3+9 k^2 M+4 k M^2+12 M\right) (k+2 N)}{12 k (k+M) (3 k+2 M)} \, , 
\quad  a_{17} = \frac{a_1 N (k+2 N)}{6 (k+M) (3 k+2 M)}  \nonumber
\end{align}
up to an overall factor $a_1$.
At $k = - 2 N$, all coefficients vanish except for $a_1,a_2$, and the charged spin 3 currents reduces to \eqref{spin3critical}.
We have checked the above expressions for $N=2,\ldots,6$ and $M=2, \ldots ,5$,
and claim that they are also true for generic $N,M$.
For $M > 3$, the condition uniquely fixes all the parameters except for $a_1$.
For $M=2,3$ we can change 
\begin{align}
a_{17} \to a_{17} - \delta \, , \quad
a_3 \to a_3 - 3 \delta\, , \quad  a_{11} \to a_{11} + 3 \delta \, , \quad a_{16} \to a_{16} + 2 M \delta \, .
\end{align}
Moreover, for $M=2$, we can use arbitrary $a_5 ,a_9 ,a_{10}$ since $d^{abc}=0$.
Other parameters can be fixed by the condition.

For the singlet currents, we need the composite operators in the trivial representation both of su$(N)$ and su$(M)$, and use the ansatz
\begin{align}
W &= a_1  d^{\alpha \beta \gamma} (J^\alpha (J ^ \beta J^\gamma ) )+ a_2 d^{abc} (J^a (J^b J^c)) + a_3 (J^{\text{u}(1)} (J^{\text{u}(1)}J^{\text{u}(1)})) \nonumber \\ &+ a_4  (J^\alpha ( J^\alpha J^{\text{u}(1)}))+ a_5 (J^a (J^a J^{\text{u}(1)})) + a_6  \left[(J^\alpha (J^{\rho \bar \imath} J^{\bar \sigma i})  +  ( J^\alpha ( J^{\bar \sigma i} J^{\rho \bar \imath} ))  \right] t^\alpha_{\rho \bar \sigma} \delta_{i \bar \imath}\nonumber  \\ &+ a_7   \left[  (J^a (J^{\rho \bar \imath} J^{\bar \rho j})) + (J^a ( J^{\bar \rho  j} J^{\rho \bar \imath} )) \right] \delta_{\rho \bar \rho} t^a_{j \bar \imath}+ a_8  \left[ (J^{\text{u}(1)} (J^{\bar \rho i} J^{\rho \bar \imath})+ (J^{\text{u}(1)} (J^{\rho \bar \imath} J^{\bar \rho i})) \right] \delta_{\rho \bar \rho}\delta_{i \bar \imath}\nonumber  \\ &+ a_9 \left( ( \partial J^\alpha )  J^\alpha \right)  + a_{10}  ((\partial J^a) J^a ) + a_{11} ( (\partial J^{\text{u}(1)}) J^{\text{u}(1)} ) \\ &+ \left[ a_{12} \left (( \partial J^{\rho \bar \imath}) J^{\bar \rho i} \right)+ a_{13} \left( ( J^{\bar \rho i}) J^{\rho \bar \imath} \right)\right] \delta_{\rho \bar \rho} \delta_{i \bar \imath}+ a_{14} \partial^2 J^{\text{u}(1)} \, .\nonumber 
\end{align}
We require that $W$ is spin 3 primary with respect to the Virasoro algebra and has the regular OPE with $ J^a $ as
\begin{align}
T (z) W (0) \sim \frac{3 W(0)}{z^2} + \frac{\partial W(0)}{z} \, ,  \quad
J^a (z) W (0)  \sim  0
\label{opeW}
\end{align}
in addition to the condition with $J^\alpha , J^{\text{u}(1)}$.
This determines the coefficients as
\begin{align}
&a_{2} = -\frac{a_1 N (k+N) (k+2 N)}{M (k+M) (k+2 M)} \, , \quad a_{3} = \frac{2 a_1 (k+N) (k+2 N) (M+N) }{k^2 M}\sqrt{\frac{M+N}{M N}} \, , \nonumber \\  
&a_{4} = \frac{6 a_1 (k+N) }{k}\sqrt{\frac{M+N}{M N}} \, , \quad a_{5} = \frac{6 a_1 (k+N) (k+2 N) }{M k (k+2 M)} \sqrt{\frac{N (M+N)}{M}} \, ,\nonumber \\  
&a_{6} = -\frac{3 a_1 (k+N)}{M} \, , \quad a_{7} = \frac{3 a_1 (k+N) (k+2 N)}{M (k+2 M)} \, , \label{Uncharged3}\\ 
&a_{8} = -\frac{3 a_1 (k+N) (k+2 N) }{k M} \sqrt{\frac{M+N}{M N}} \, , \quad a_{9} =  0 \, , \quad a_{10} =  0  \, , \quad a_{11} =  0\, , \nonumber \\  
&a_{12} = \frac{3 a_1 (k+N) (k+2 N)}{M} \, , \quad a_{13} =  -\frac{3 a_1 (k+N) (k+2 N)}{M} \, , \nonumber \\
&a_{14} = - a_1 (k+N) (k+2 N) \sqrt{\frac{N (M+N)}{M}}  \, .\nonumber 
\end{align}
We have obtained the above expressions with $N=2,\ldots ,6$ and $M=2,\ldots ,5$, and propose that they are true also for generic $N,M$.
For $N , M >2$, these coefficients are uniquely fixed up to an overall normalization $a_1$.
For $N=2$, there is no relation between $a_1$ and other coefficients because of $d^{\alpha \beta \gamma} = 0$.
Similarly, for $M=2$, we can use arbitrary $a_2$ due to $ d^{a b c} = 0$.
Utilizing these fact, the coefficients can be chosen as in \eqref{Uncharged3}, and others are fixed uniquely. 
At $k= - 2 N$, the expressions become simplified and only $a_1,a_4,a_6$ remain non-zero.
In this case, we have checked for several examples%
\footnote{We have examined the cases with $N,M = 2,3,4$.}
that $W (z) W (0) \sim 0 \cdot z^{-6} + \mathcal{O}(z^{-5})$. 
Therefore, we can say that the spin 3 current $W$ is decoupled  at $k= - 2 N$ just as in the case of $P^a$.

\section{Decomposition of  W-algebra with $(n,M)=(2,2)$}
\label{sec:webofwalg}

In this appendix, we reproduce the decomposition in \eqref{cosetdec} from the W$_2$-algebra with $\text{su}(2)$ symmetry obtained in subsection  \ref{sec:extVirasoro}. Similar analysis for other type of W-algebras has been done in \cite{Prochazka:2017qum}, see also \cite{Gaberdiel:2017hcn}.

First, we look for two commuting energy-momentum tensors $T_\eta$ $(\eta=0,1)$, whose OPEs are
\begin{align}
T_\eta (z) T_\eta (0) \sim \frac{C_i/2}{z^4} + \frac{2 T_\eta}{z^2} + \frac{\partial T_\eta}{z} \, , \quad
T_0 (z) T_1 (0) \sim 0 \, , \quad  T_\eta (z) J^3 (0) \sim 0 \, .
\label{opedec}
\end{align}
Here we have chosen $t^3$ as the generator of Cartan sub-algebra and  required that the spin 1 current $J^3$ decouples from the algebras.
We use the ansatz
\begin{align}
\begin{aligned}
&T_0 \equiv a_1 T  + a_2  Q^3 + a_3 (J^a J^a) + a_4 (J^3 J^3) \, , \\
&T_1 \equiv b_1 T  + b_2  Q^3 + b_3 (J^a J^a) + b_4 (J^3 J^3) \, .
\end{aligned}
\end{align}
Then the OPEs in \eqref{opedec} are reproduced with
\begin{align}
\begin{aligned}
&a_1 =  \frac{1}{k+2}+\frac{1}{2} \, , \quad a_2 =  \frac{\sqrt{- 3 k^3 - k^2 + 4 k }}{\sqrt{2} (k+2)}\, , \quad a_3 =  -\frac{1}{2 k+4} \, , \quad a_4 = \frac{1}{4 (k+2)} \, ,\\
&b_1 =  \frac{k}{2 k+4} \, , \quad b_2 =  -\frac{\sqrt{-3 k^3-k^2+4 k}}{\sqrt{2} (k+2)} \, , \quad b_3 =  \frac{1}{2 k+4} \, , \quad b_4 = -\frac{3 k+4}{4 k^2+8 k} \, .
\end{aligned}
\end{align}
The central charges are $C_0 = c_{k,N}$ and $C_1 = c_{k,N+1}$ as they should be.

Next, we search for spin 3 and 4 currents, which are primary with respect to $T_0$ or $T_1$ and commute with $J^3$.
There are no such currents primary to $T_0$. The three parameters $\mu_0^{(i)}$ $(i=1,2,3)$ for the W-algebra including $T_0$ are
\begin{align}
\mu_1^{(0)} = 2 \, , \quad \mu_2^{(0)} = - \frac{2 k}{k +2} \, ,  \quad  \mu_3^{(0)} = k \,  .
\end{align}
This implies that the allowed spin of current is only $2$, and this is consistent with our result. We can find out spin 3 and 4 currents primary to $T_1$. We normalize the currents such as
\begin{align}
W^{(3)} =  (J^a Q^a) + \cdots \, , \qquad W^{(4)} = (Q^a Q^a) + \cdots \, .
\end{align}
Some of the coefficients of OPEs are
\begin{align}
\begin{aligned}
&C^{0}_{33} =  \frac{2 (k-2)^2 (k+4)}{3 k^2} \, , \quad
C^4_{33} = k + 4 \, , \\
&C^0_{44}  =  -\frac{4 (k-4) (k-3) (k-2)^2 (k+1) (k+4) (k+6) (3 k+2) (5 k+12)}{(k-1) k^4 (3 k+4) (k (k (15 k-47)-162)-136)} \, ,
\end{aligned}
\end{align}
where $C_{ij}^k$ are the coefficient in front of $W^{(k)}$ in the OPE of $W^{(i)}$ and W$^{(j)}$.
According to \cite{Prochazka:2014gqa} (see also \cite{Gaberdiel:2012ku}), their normalization invariant combination is related to the parameters $\lambda_i$ $(i=1,2,3)$ and the central charge $c$ of the W-algebra as
\begin{align}
\frac{(C_{33}^4)^2 C_{44}^0}{(C_{33}^0)^2} = \frac{144 (c+2) (\lambda_1 -3) (\lambda_2 -3) (\lambda_3 -3)}{ c (5 c+22) (\lambda_1 -2) (\lambda_2 -2) (\lambda_3 -2)} \, .
\end{align}
This equation is consistent with the values $\lambda_i = \mu^{(1)}_i$ and $c = c_{k , N + 1}$ if we set $k/(k + N) = 2$.


\begin{thebibliography}{10}
	
	\bibitem{Gross:1988ue}
	D.~J. Gross, \emph{{High-energy symmetries of string theory}},
	\href{https://doi.org/10.1103/PhysRevLett.60.1229}{\emph{Phys.Rev.Lett.}
		{\bfseries 60} (1988) 1229}.
	
	\bibitem{Vasiliev:2003ev}
	M.~Vasiliev, \emph{{Nonlinear equations for symmetric massless higher spin
			fields in (A)dS$_{(d)}$}},
	\href{https://doi.org/10.1016/S0370-2693(03)00872-4}{\emph{Phys.Lett.}
		{\bfseries B567} (2003) 139}
	[\href{https://arxiv.org/abs/hep-th/0304049}{{\ttfamily hep-th/0304049}}].
	
	\bibitem{Vasiliev:2018zer}
	M.~A. Vasiliev, \emph{{From Coxeter higher-spin theories to strings and tensor
			models}}, \href{https://doi.org/10.1007/JHEP08(2018)051}{\emph{JHEP}
		{\bfseries 08} (2018) 051}
	[\href{https://arxiv.org/abs/1804.06520}{{\ttfamily 1804.06520}}].
	
	\bibitem{Chang:2012kt}
	C.-M. Chang, S.~Minwalla, T.~Sharma and X.~Yin, \emph{{ABJ triality: From
			higher spin fields to strings}},
	\href{https://doi.org/10.1088/1751-8113/46/21/214009}{\emph{J.Phys.}
		{\bfseries A46} (2013) 214009}
	[\href{https://arxiv.org/abs/1207.4485}{{\ttfamily 1207.4485}}].
	
	\bibitem{Aharony:2008ug}
	O.~Aharony, O.~Bergman, D.~L. Jafferis and J.~Maldacena, \emph{{$\mathcal{N}=6$
			superconformal Chern-Simons-matter theories, M2-branes and their gravity
			duals}}, \href{https://doi.org/10.1088/1126-6708/2008/10/091}{\emph{JHEP}
		{\bfseries 0810} (2008) 091}
	[\href{https://arxiv.org/abs/0806.1218}{{\ttfamily 0806.1218}}].
	
	\bibitem{Aharony:2008gk}
	O.~Aharony, O.~Bergman and D.~L. Jafferis, \emph{{Fractional M2-branes}},
	\href{https://doi.org/10.1088/1126-6708/2008/11/043}{\emph{JHEP} {\bfseries
			0811} (2008) 043} [\href{https://arxiv.org/abs/0807.4924}{{\ttfamily
			0807.4924}}].
	
	\bibitem{Creutzig:2013tja}
	T.~Creutzig, Y.~Hikida and P.~B. R{\o}nne, \emph{{Extended higher spin
			holography and Grassmannian models}},
	\href{https://doi.org/10.1007/JHEP11(2013)038}{\emph{JHEP} {\bfseries 1311}
		(2013) 038} [\href{https://arxiv.org/abs/1306.0466}{{\ttfamily 1306.0466}}].
	
	\bibitem{Prokushkin:1998bq}
	S.~Prokushkin and M.~A. Vasiliev, \emph{{Higher spin gauge interactions for
			massive matter fields in 3-D AdS space-time}},
	\href{https://doi.org/10.1016/S0550-3213(98)00839-6}{\emph{Nucl.Phys.}
		{\bfseries B545} (1999) 385}
	[\href{https://arxiv.org/abs/hep-th/9806236}{{\ttfamily hep-th/9806236}}].
	
	\bibitem{Eberhardt:2018plx}
	L.~Eberhardt, M.~R. Gaberdiel and I.~Rienacker, \emph{{Higher spin algebras and
			large $ \mathcal{N} $ = 4 holography}},
	\href{https://doi.org/10.1007/JHEP03(2018)097}{\emph{JHEP} {\bfseries 03}
		(2018) 097} [\href{https://arxiv.org/abs/1801.00806}{{\ttfamily
			1801.00806}}].
	
	\bibitem{Gaberdiel:2010pz}
	M.~R. Gaberdiel and R.~Gopakumar, \emph{{An AdS$_3$ dual for minimal model
			CFTs}}, \href{https://doi.org/10.1103/PhysRevD.83.066007}{\emph{Phys.Rev.}
		{\bfseries D83} (2011) 066007}
	[\href{https://arxiv.org/abs/1011.2986}{{\ttfamily 1011.2986}}].
	
	\bibitem{Creutzig:2011fe}
	T.~Creutzig, Y.~Hikida and P.~B. R{\o}nne, \emph{{Higher spin AdS$_3$
			supergravity and its dual CFT}},
	\href{https://doi.org/10.1007/JHEP02(2012)109}{\emph{JHEP} {\bfseries 1202}
		(2012) 109} [\href{https://arxiv.org/abs/1111.2139}{{\ttfamily 1111.2139}}].
	
	\bibitem{Joung:2017hsi}
	E.~Joung, J.~Kim, J.~Kim and S.-J. Rey, \emph{{Asymptotic symmetries of colored
			gravity in three dimensions}},
	\href{https://doi.org/10.1007/JHEP03(2018)104}{\emph{JHEP} {\bfseries 03}
		(2018) 104} [\href{https://arxiv.org/abs/1712.07744}{{\ttfamily
			1712.07744}}].
	
	\bibitem{Gwak:2015vfb}
	S.~Gwak, E.~Joung, K.~Mkrtchyan and S.-J. Rey, \emph{{Rainbow valley of colored
			(anti-)de Sitter gravity in three dimensions}},
	\href{https://doi.org/10.1007/JHEP04(2016)055}{\emph{JHEP} {\bfseries 04}
		(2016) 055} [\href{https://arxiv.org/abs/1511.05220}{{\ttfamily
			1511.05220}}].
	
	\bibitem{Gwak:2015jdo}
	S.~Gwak, E.~Joung, K.~Mkrtchyan and S.-J. Rey, \emph{{Rainbow vacua of colored
			higher spin gravity in three dimensions}},
	\href{https://arxiv.org/abs/1511.05975}{{\ttfamily 1511.05975}}.
	
	\bibitem{Henneaux:2010xg}
	M.~Henneaux and S.-J. Rey, \emph{{Nonlinear $W_\infty$ as asymptotic symmetry
			of three-dimensional higher spin anti-de Sitter gravity}},
	\href{https://doi.org/10.1007/JHEP12(2010)007}{\emph{JHEP} {\bfseries 1012}
		(2010) 007} [\href{https://arxiv.org/abs/1008.4579}{{\ttfamily 1008.4579}}].
	
	\bibitem{Campoleoni:2010zq}
	A.~Campoleoni, S.~Fredenhagen, S.~Pfenninger and S.~Theisen, \emph{{Asymptotic
			symmetries of three-dimensional gravity coupled to higher-spin fields}},
	\href{https://doi.org/10.1007/JHEP11(2010)007}{\emph{JHEP} {\bfseries 1011}
		(2010) 007} [\href{https://arxiv.org/abs/1008.4744}{{\ttfamily 1008.4744}}].
	
	\bibitem{Gaberdiel:2011wb}
	M.~R. Gaberdiel and T.~Hartman, \emph{{Symmetries of holographic minimal
			models}}, \href{https://doi.org/10.1007/JHEP05(2011)031}{\emph{JHEP}
		{\bfseries 1105} (2011) 031}
	[\href{https://arxiv.org/abs/1101.2910}{{\ttfamily 1101.2910}}].
	
	\bibitem{Campoleoni:2011hg}
	A.~Campoleoni, S.~Fredenhagen and S.~Pfenninger, \emph{{Asymptotic W-symmetries
			in three-dimensional higher-spin gauge theories}},
	\href{https://doi.org/10.1007/JHEP09(2011)113}{\emph{JHEP} {\bfseries 1109}
		(2011) 113} [\href{https://arxiv.org/abs/1107.0290}{{\ttfamily 1107.0290}}].
	
	\bibitem{Gaberdiel:2012ku}
	M.~R. Gaberdiel and R.~Gopakumar, \emph{{Triality in minimal model
			holography}}, \href{https://doi.org/10.1007/JHEP07(2012)127}{\emph{JHEP}
		{\bfseries 07} (2012) 127} [\href{https://arxiv.org/abs/1205.2472}{{\ttfamily
			1205.2472}}].
	
	\bibitem{Prochazka:2014gqa}
	T.~Procházka, \emph{{Exploring $ {\mathcal{W}}_{\infty } $ in the quadratic
			basis}}, \href{https://doi.org/10.1007/JHEP09(2015)116}{\emph{JHEP}
		{\bfseries 09} (2015) 116} [\href{https://arxiv.org/abs/1411.7697}{{\ttfamily
			1411.7697}}].
	
	\bibitem{Candu:2012tr}
	C.~Candu and M.~R. Gaberdiel, \emph{{Duality in $\mathcal{N}=2$ minimal model
			holography}}, \href{https://doi.org/10.1007/JHEP02(2013)070}{\emph{JHEP}
		{\bfseries 02} (2013) 070} [\href{https://arxiv.org/abs/1207.6646}{{\ttfamily
			1207.6646}}].
	
	\bibitem{Gaiotto:2017euk}
	D.~Gaiotto and M.~Rapčák, \emph{{Vertex algebras at the corner}},
	\href{https://arxiv.org/abs/1703.00982}{{\ttfamily 1703.00982}}.
	
	\bibitem{Creutzig:2017uxh}
	T.~Creutzig and D.~Gaiotto, \emph{{Vertex algebras for S-duality}},
	\href{https://arxiv.org/abs/1708.00875}{{\ttfamily 1708.00875}}.
	
	\bibitem{Prochazka:2017qum}
	T.~Procházka and M.~Rapčák, \emph{{Webs of W-algebras}},
	\href{https://arxiv.org/abs/1711.06888}{{\ttfamily 1711.06888}}.
	
	\bibitem{Prochazka:2018tlo}
	T.~Procházka and M.~Rapčák, \emph{{$\mathcal{W}$-algebra modules, free
			fields, and Gukov-Witten defects}},
	\href{https://arxiv.org/abs/1808.08837}{{\ttfamily 1808.08837}}.
	
	\bibitem{Harada:2018bkb}
	K.~Harada and Y.~Matsuo, \emph{{Plane partition realization of (web of)
			W-algebra minimal models}},
	\href{https://arxiv.org/abs/1810.08512}{{\ttfamily 1810.08512}}.
	
	\bibitem{Achucarro:1987vz}
	A.~Achucarro and P.~Townsend, \emph{{A Chern-Simons action for
			three-dimensional anti-de Sitter supergravity theories}},
	\href{https://doi.org/10.1016/0370-2693(86)90140-1}{\emph{Phys.Lett.}
		{\bfseries B180} (1986) 89}.
	
	\bibitem{Witten:1988hc}
	E.~Witten, \emph{{$(2+1)-$dimensional gravity as an exactly soluble system}},
	\href{https://doi.org/10.1016/0550-3213(88)90143-5}{\emph{Nucl.Phys.}
		{\bfseries B311} (1988) 46}.
	
	\bibitem{Blencowe:1988gj}
	M.~Blencowe, \emph{{A consistent interacting massless higher spin field theory
			in $D = (2+1)$}},
	\href{https://doi.org/10.1088/0264-9381/6/4/005}{\emph{Class.Quant.Grav.}
		{\bfseries 6} (1989) 443}.
	
	\bibitem{Pope:1989sr}
	C.~N. Pope, L.~J. Romans and X.~Shen, \emph{{W$_\infty$ and the Racah-Wigner
			algebra}}, \href{https://doi.org/10.1016/0550-3213(90)90539-P}{\emph{Nucl.
			Phys.} {\bfseries B339} (1990) 191}.
	
	\bibitem{Gaberdiel:2013vva}
	M.~R. Gaberdiel and R.~Gopakumar, \emph{{Large $\mathcal{N}=4$ holography}},
	\href{https://doi.org/10.1007/JHEP09(2013)036}{\emph{JHEP} {\bfseries 1309}
		(2013) 036} [\href{https://arxiv.org/abs/1305.4181}{{\ttfamily 1305.4181}}].
	
	\bibitem{Castro:2011iw}
	A.~Castro, R.~Gopakumar, M.~Gutperle and J.~Raeymaekers, \emph{{Conical defects
			in higher spin theories}},
	\href{https://doi.org/10.1007/JHEP02(2012)096}{\emph{JHEP} {\bfseries 02}
		(2012) 096} [\href{https://arxiv.org/abs/1111.3381}{{\ttfamily 1111.3381}}].
	
	\bibitem{Balog:1990mu}
	J.~Balog, L.~Feher, L.~O'Raifeartaigh, P.~Forgacs and A.~Wipf, \emph{{Toda
			theory and $W$ algebra from a gauged {WZNW} point of view}},
	\href{https://doi.org/10.1016/0003-4916(90)90029-N}{\emph{Annals Phys.}
		{\bfseries 203} (1990) 76}.
	
	\bibitem{Campoleoni:2017xyl}
	A.~Campoleoni, S.~Fredenhagen and J.~Raeymaekers, \emph{{Quantizing higher-spin
			gravity in free-field variables}},
	\href{https://doi.org/10.1007/JHEP02(2018)126}{\emph{JHEP} {\bfseries 02}
		(2018) 126} [\href{https://arxiv.org/abs/1712.08078}{{\ttfamily
			1712.08078}}].
	
	\bibitem{Henneaux:1999ib}
	M.~Henneaux, L.~Maoz and A.~Schwimmer, \emph{{Asymptotic dynamics and
			asymptotic symmetries of three-dimensional extended AdS supergravity}},
	\href{https://doi.org/10.1006/aphy.2000.5994}{\emph{Annals Phys.} {\bfseries
			282} (2000) 31} [\href{https://arxiv.org/abs/hep-th/9910013}{{\ttfamily
			hep-th/9910013}}].
	
	\bibitem{Arakawa:2016rwm}
	T.~Arakawa, \emph{{Introduction to W-algebras and their representation
			theory}},  \href{https://arxiv.org/abs/1605.00138}{{\ttfamily 1605.00138}}.
	
	\bibitem{Creutzig:2012sf}
	T.~Creutzig and A.~R. Linshaw, \emph{{The super W$_{1+\infty}$ algebra with
			integral central charge}},
	\href{https://doi.org/10.1090/S0002-9947-2015-06214-X}{\emph{Trans. Am. Math.
			Soc.} {\bfseries 367} (2015) 5521}
	[\href{https://arxiv.org/abs/1209.6032}{{\ttfamily 1209.6032}}].
	
	\bibitem{Creutzig:2014lsa}
	T.~Creutzig and A.~R. Linshaw, \emph{{Cosets of affine vertex algebras inside
			larger structures}},
	\href{https://doi.org/10.1016/j.jalgebra.2018.10.007}{\emph{J. Algebra}
		{\bfseries 517} (2019) 396}
	[\href{https://arxiv.org/abs/1407.8512}{{\ttfamily 1407.8512}}].
	
	\bibitem{Kac:2003jh}
	V.~G. Kac and M.~Wakimoto, \emph{{Quantum reduction and representation theory
			of superconformal algebras}},
	\href{https://arxiv.org/abs/math-ph/0304011}{{\ttfamily math-ph/0304011}}.
	
	\bibitem{Genra:2016xxc}
	N.~Genra, \emph{{Screening operators for W-algebras}},
	\href{https://arxiv.org/abs/1606.00966}{{\ttfamily 1606.00966}}.
	
	\bibitem{Arakawa:2018iyk}
	T.~Arakawa, T.~Creutzig and A.~R. Linshaw, \emph{{W-algebras as coset vertex
			algebras}},  \href{https://arxiv.org/abs/1801.03822}{{\ttfamily 1801.03822}}.
	
	\bibitem{Linshaw:2017tvv}
	A.~R. Linshaw, \emph{{Universal two-parameter $\mathcal{W}_{\infty}$-algebra
			and vertex algebras of type $\mathcal{W}(2,3,\dots, N)$}},
	\href{https://arxiv.org/abs/1710.02275}{{\ttfamily 1710.02275}}.
	
	\bibitem{Kanade:2018qut}
	S.~Kanade and A.~R. Linshaw, \emph{{Universal two-parameter even spin
			$\mathcal{W}_{\infty}$-algebra}},
	\href{https://arxiv.org/abs/1805.11031}{{\ttfamily 1805.11031}}.
	
	\bibitem{Arakawa:2016fbi}
	T.~Arakawa and A.~Molev, \emph{{Explicit generators in rectangular affine
			$\mathcal {W}$ -algebras of type A}},
	\href{https://doi.org/10.1007/s11005-016-0890-2}{\emph{Lett. Math. Phys.}
		{\bfseries 107} (2017) 47} [\href{https://arxiv.org/abs/1403.1017}{{\ttfamily
			1403.1017}}].
	
	\bibitem{Creutzig:2015hla}
	T.~Creutzig, Y.~Hikida and P.~B. Rønne, \emph{{Correspondences between WZNW
			models and CFTs with W-algebra symmetry}},
	\href{https://doi.org/10.1007/JHEP02(2016)048}{\emph{JHEP} {\bfseries 02}
		(2016) 048} [\href{https://arxiv.org/abs/1509.07516}{{\ttfamily
			1509.07516}}].
	
	\bibitem{Hikida:2007tq}
	Y.~Hikida and V.~Schomerus, \emph{{$H_3^+$ WZNW model from Liouville field
			theory}}, \href{https://doi.org/10.1088/1126-6708/2007/10/064}{\emph{JHEP}
		{\bfseries 10} (2007) 064} [\href{https://arxiv.org/abs/0706.1030}{{\ttfamily
			0706.1030}}].
	
	\bibitem{Hikida:2007sz}
	Y.~Hikida and V.~Schomerus, \emph{{Structure constants of the OSP$(1|2)$ WZNW
			model}}, \href{https://doi.org/10.1088/1126-6708/2007/12/100}{\emph{JHEP}
		{\bfseries 12} (2007) 100} [\href{https://arxiv.org/abs/0711.0338}{{\ttfamily
			0711.0338}}].
	
	\bibitem{Creutzig:2011qm}
	T.~Creutzig, Y.~Hikida and P.~B. Rønne, \emph{{Supergroup - extended super
			Liouville correspondence}},
	\href{https://doi.org/10.1007/JHEP06(2011)063}{\emph{JHEP} {\bfseries 06}
		(2011) 063} [\href{https://arxiv.org/abs/1103.5753}{{\ttfamily 1103.5753}}].
	
	\bibitem{Bakas:1990xu}
	I.~Bakas and E.~Kiritsis, \emph{{Grassmannian coset models and unitary
			representations of W$_\infty$}},
	\href{https://doi.org/10.1142/S0217732390002328}{\emph{Mod. Phys. Lett.}
		{\bfseries A5} (1990) 2039}.
	
	\bibitem{Odake:1990rr}
	S.~Odake and T.~Sano, \emph{{W$_{1 + \infty}$ and super W$_\infty$ algebras
			with SU$(N)$ symmetry}},
	\href{https://doi.org/10.1016/0370-2693(91)91101-Z}{\emph{Phys. Lett.}
		{\bfseries B258} (1991) 369}.
	
	\bibitem{Thielemans:1991uw}
	K.~Thielemans, \emph{{A Mathematica package for computing operator product
			expansions}}, \href{https://doi.org/10.1142/S0129183191001001}{\emph{Int. J.
			Mod. Phys.} {\bfseries C2} (1991) 787}.
	
	\bibitem{DiFrancesco:1997nk}
	P.~Di~Francesco, P.~Mathieu and D.~Senechal, \emph{{Conformal field theory}},
	Graduate Texts in Contemporary Physics. Springer-Verlag, New York, 1997,
	\href{https://doi.org/10.1007/978-1-4612-2256-9}{10.1007/978-1-4612-2256-9}.
	
	\bibitem{Creutzig:2016ehb}
	T.~Creutzig, S.~Kanade, A.~R. Linshaw and D.~Ridout, \emph{{Schur-Weyl duality
			for Heisenberg cosets}},
	\href{https://doi.org/10.1007/s00031-018-9497-2}{\emph{Transformation Groups}
		(2018) 1} [\href{https://arxiv.org/abs/1611.00305}{{\ttfamily 1611.00305}}].
	
	\bibitem{Creutzig:2017qyf}
	T.~Creutzig, \emph{{W-algebras for Argyres-Douglas theories}},
	\href{https://doi.org/10.1007/s40879-017-0156-2}{\emph{European Journal of
			Mathematics} {\bfseries 3} (2017) 659}
	[\href{https://arxiv.org/abs/1701.05926}{{\ttfamily 1701.05926}}].
	
	\bibitem{2018arXiv180509771A}
	D.~{Adamovic} and A.~{Milas}, \emph{{On some vertex algebras related to
			$V_{-1}(\frak{sl} (n) )$ and their characters}},
	\href{https://arxiv.org/abs/1805.09771}{{\ttfamily 1805.09771}}.
	
	\bibitem{KWbdy}
	V.~G. {Kac} and M.~{Wakimoto}, \emph{{A remark on boundary level admissible
			representations}},  \href{https://arxiv.org/abs/1612.07423}{{\ttfamily
			1612.07423}}.
	
	\bibitem{Creutzig:2013hma}
	T.~Creutzig and D.~Ridout, \emph{{Logarithmic Conformal Field Theory: Beyond an
			Introduction}},
	\href{https://doi.org/10.1088/1751-8113/46/49/494006}{\emph{J. Phys.}
		{\bfseries A46} (2013) 4006}
	[\href{https://arxiv.org/abs/1303.0847}{{\ttfamily 1303.0847}}].
	
	\bibitem{MR1359963}
	D.~Adamovi\'{c} and A.~Milas, \emph{Vertex operator algebras associated to
		modular invariant representations for {$A^{(1)}_1$}},
	\href{https://doi.org/10.4310/MRL.1995.v2.n5.a4}{\emph{Math. Res. Lett.}
		{\bfseries 2} (1995) 563}.
	
	\bibitem{MR3093193}
	T.~Creutzig and D.~Ridout, \emph{Modular data and {V}erlinde formulae for
		fractional level {WZW} models {II}},
	\href{https://doi.org/10.1016/j.nuclphysb.2013.07.008}{\emph{Nuclear Phys. B}
		{\bfseries 875} (2013) 423}.
	
	\bibitem{MR3333645}
	D.~Ridout and S.~Wood, \emph{Relaxed singular vectors, {J}ack symmetric
		functions and fractional level {$\widehat{{sl}}(2)$} models},
	\href{https://doi.org/10.1016/j.nuclphysb.2015.03.023}{\emph{Nuclear Phys. B}
		{\bfseries 894} (2015) 621}.
	
	\bibitem{MR1026952}
	V.~G. Kac and M.~Wakimoto, \emph{Classification of modular invariant
		representations of affine algebras},  in \emph{Infinite-dimensional {L}ie
		algebras and groups ({L}uminy-{M}arseille, 1988)}, vol.~7 of \emph{Adv. Ser.
		Math. Phys.}, pp.~138--177.
	\newblock World Sci. Publ., Teaneck, NJ, 1989.
	
	\bibitem{arakawa2016}
	T.~Arakawa, \emph{Rationality of admissible affine vertex algebras in the
		category ${{O}}$}, \href{https://doi.org/10.1215/00127094-3165113}{\emph{Duke
			Math. J.} {\bfseries 165} (2016) 67}.
	
	\bibitem{MR949675}
	V.~G. Kac and M.~Wakimoto, \emph{Modular invariant representations of
		infinite-dimensional {L}ie algebras and superalgebras},
	\href{https://doi.org/10.1073/pnas.85.14.4956}{\emph{Proc. Nat. Acad. Sci.
			U.S.A.} {\bfseries 85} (1988) 4956}.
	
	\bibitem{MR3845289}
	T.~Creutzig, Y.-Z. Huang and J.~Yang, \emph{Braided tensor categories of
		admissible modules for affine {L}ie algebras},
	\href{https://doi.org/10.1007/s00220-018-3217-6}{\emph{Comm. Math. Phys.}
		{\bfseries 362} (2018) 827}.
	
	\bibitem{2018arXiv180700415C}
	T.~{Creutzig}, \emph{{Fusion categories for affine vertex algebras at
			admissible levels}},  \href{https://arxiv.org/abs/1807.00415}{{\ttfamily
			1807.00415}}.
	
	\bibitem{BCR14}
	K.~Bringmann, T.~Creutzig and L.~Rolen, \emph{Negative index jacobi forms and
		quantum modular forms},
	\href{https://doi.org/10.1186/s40687-014-0011-8}{\emph{Research in the
			Mathematical Sciences} {\bfseries 1} (2014) 11}.
	
	\bibitem{BRZ16}
	K.~Bringmann, L.~Rolen and S.~Zwegers, \emph{On the fourier coefficients of
		negative index meromorphic jacobi forms},
	\href{https://doi.org/10.1186/s40687-016-0056-y}{\emph{Research in the
			Mathematical Sciences} {\bfseries 3} (2016) 5}.
	
	\bibitem{Candu:2013fta}
	C.~Candu and C.~Vollenweider, \emph{{On the coset duals of extended higher spin
			theories}}, \href{https://doi.org/10.1007/JHEP04(2014)145}{\emph{JHEP}
		{\bfseries 04} (2014) 145} [\href{https://arxiv.org/abs/1312.5240}{{\ttfamily
			1312.5240}}].
	
	\bibitem{Creutzig:2014ula}
	T.~Creutzig, Y.~Hikida and P.~B. R{\o}nne, \emph{{Higher spin AdS$_{3}$
			holography with extended supersymmetry}},
	\href{https://doi.org/10.1007/JHEP10(2014)163}{\emph{JHEP} {\bfseries 1410}
		(2014) 163} [\href{https://arxiv.org/abs/1406.1521}{{\ttfamily 1406.1521}}].
	
	\bibitem{Bais:1987dc}
	F.~A. Bais, P.~Bouwknegt, M.~Surridge and K.~Schoutens, \emph{{Extensions of
			the Virasoro algebra constructed from Kac-Moody algebras using higher order
			casimir invariants}},
	\href{https://doi.org/10.1016/0550-3213(88)90631-1}{\emph{Nucl. Phys.}
		{\bfseries B304} (1988) 348}.
	
	\bibitem{Tsymbaliuk}
	O.~Tsymbaliuk, \emph{{The affine Yangian of gl$_1$, and the infnitesimal
			Cherednik algebras}}, {\emph{Ph.D. thesis, MIT, Department of Mathematics}
		(2014) }.
	
	\bibitem{Prochazka:2015deb}
	T.~Procházka, \emph{{$ \mathcal{W} $-symmetry, topological vertex and affine
			Yangian}}, \href{https://doi.org/10.1007/JHEP10(2016)077}{\emph{JHEP}
		{\bfseries 10} (2016) 077}
	[\href{https://arxiv.org/abs/1512.07178}{{\ttfamily 1512.07178}}].
	
	\bibitem{Gaberdiel:2017dbk}
	M.~R. Gaberdiel, R.~Gopakumar, W.~Li and C.~Peng, \emph{{Higher spins and
			Yangian symmetries}},
	\href{https://doi.org/10.1007/JHEP04(2017)152}{\emph{JHEP} {\bfseries 04}
		(2017) 152} [\href{https://arxiv.org/abs/1702.05100}{{\ttfamily
			1702.05100}}].
	
	\bibitem{Altschuler:1989nm}
	D.~Altschuler, M.~Bauer and C.~Itzykson, \emph{{The branching rules of
			conformal embeddings}},
	\href{https://doi.org/10.1007/BF02096653}{\emph{Commun. Math. Phys.}
		{\bfseries 132} (1990) 349}.
	
	\bibitem{Walton:1988bs}
	M.~A. Walton, \emph{{Conformal branching rules and modular invariants}},
	\href{https://doi.org/10.1016/0550-3213(89)90237-X}{\emph{Nucl. Phys.}
		{\bfseries B322} (1989) 775}.
	
	\bibitem{Gaberdiel:2017hcn}
	M.~R. Gaberdiel, W.~Li, C.~Peng and H.~Zhang, \emph{{The supersymmetric affine
			Yangian}}, \href{https://doi.org/10.1007/JHEP05(2018)200}{\emph{JHEP}
		{\bfseries 05} (2018) 200}
	[\href{https://arxiv.org/abs/1711.07449}{{\ttfamily 1711.07449}}].
	
	\bibitem{Gaberdiel:2018nbs}
	M.~R. Gaberdiel, W.~Li and C.~Peng, \emph{{Twin-plane-partitions and
			$\mathcal{N}=2$ affine Yangian}},
	\href{https://arxiv.org/abs/1807.11304}{{\ttfamily 1807.11304}}.
	
	\bibitem{Perlmutter:2012ds}
	E.~Perlmutter, T.~Prochazka and J.~Raeymaekers, \emph{{The semiclassical limit
			of W$_N$ CFTs and Vasiliev theory}},
	\href{https://doi.org/10.1007/JHEP05(2013)007}{\emph{JHEP} {\bfseries 05}
		(2013) 007} [\href{https://arxiv.org/abs/1210.8452}{{\ttfamily 1210.8452}}].
	
	\bibitem{Hikida:2012eu}
	Y.~Hikida, \emph{{Conical defects and $\mathcal{N}=2$ higher spin holography}},
	\href{https://doi.org/10.1007/JHEP08(2013)127}{\emph{JHEP} {\bfseries 08}
		(2013) 127} [\href{https://arxiv.org/abs/1212.4124}{{\ttfamily 1212.4124}}].
	
	\bibitem{Hikida:2015nfa}
	Y.~Hikida and P.~B. R{\o}nne, \emph{{Marginal deformations and the Higgs
			phenomenon in higher spin AdS$_{3}$ holography}},
	\href{https://doi.org/10.1007/JHEP07(2015)125}{\emph{JHEP} {\bfseries 07}
		(2015) 125} [\href{https://arxiv.org/abs/1503.03870}{{\ttfamily
			1503.03870}}].
	
	\bibitem{Henneaux:2012ny}
	M.~Henneaux, G.~Lucena~G\'{o}mez, J.~Park and S.-J. Rey,
	\emph{{Super-$W_\infty$ asymptotic symmetry of higher-spin AdS$_3$
			supergravity}}, \href{https://doi.org/10.1007/JHEP06(2012)037}{\emph{JHEP}
		{\bfseries 1206} (2012) 037}
	[\href{https://arxiv.org/abs/1203.5152}{{\ttfamily 1203.5152}}].
	
	\bibitem{Gaberdiel:2014cha}
	M.~R. Gaberdiel and R.~Gopakumar, \emph{{Higher spins \& strings}},
	\href{https://doi.org/10.1007/JHEP11(2014)044}{\emph{JHEP} {\bfseries 1411}
		(2014) 044} [\href{https://arxiv.org/abs/1406.6103}{{\ttfamily 1406.6103}}].
	
	\bibitem{Gaberdiel:2015mra}
	M.~R. Gaberdiel and R.~Gopakumar, \emph{{Stringy symmetries and the higher spin
			square}}, \href{https://doi.org/10.1088/1751-8113/48/18/185402}{\emph{J.
			Phys.} {\bfseries A48} (2015) 185402}
	[\href{https://arxiv.org/abs/1501.07236}{{\ttfamily 1501.07236}}].
	
\end{thebibliography}

\providecommand{\href}[2]{#2}\begingroup\raggedright\endgroup

\end{document}